\documentclass[useAMS,letterpaper,usenatbib]{mn2e}
\input psfig.sty

\def \xoff {\ifmmode x_{\rm off} \else $x_{\rm off}$ \fi}
\def \rhorms {\ifmmode \rho_{\rm rms} \else $\rho_{\rm rms}$ \fi}

\def \apj  {ApJ}
\def \apjl  {ApJL}
\def \apjs  {ApJS}

\def \mnras {MNRAS}
\def \etal {et~al.~}
\def \chisq  {\ifmmode  \chi^2   \else  $\chi^2$  \fi}  
\def \chisqr {\ifmmode \chi^2_{\rm r} \else $\chi^2_{\rm r}$ \fi}
\def \spose#1{\hbox  to 0pt{#1\hss}}  
\def \lta{\mathrel{\spose{\lower 3pt\hbox{$\sim$}}\raise  2.0pt\hbox{$<$}}}
\def \gta{\mathrel{\spose{\lower  3pt\hbox{$\sim$}}\raise 2.0pt\hbox{$>$}}}
 
\def \ha  {\ifmmode H\alpha \else H$\alpha $ \fi}

\def \morgana {{\sc morgana}}

\def \kms {\ifmmode  \,\rm km\,s^{-1} \else $\,\rm km\,s^{-1}  $ \fi }
\def \kpc {\ifmmode  {\rm kpc}  \else ${\rm  kpc}$ \fi  }  
\def \Msun {\ifmmode M_{\odot} \else $M_{\odot}$ \fi} 
\def \hMsun {\ifmmode h^{-1}\,\rm M_{\odot} \else $h^{-1}\,\rm M_{\odot}$ \fi}
\def \hhMsun {\ifmmode h^{-2}\,\rm M_{\odot}\else $h^{-2}\,\rm M_{\odot}$ \fi}
\def \Lsun {\ifmmode L_{\odot} \else $L_{\odot}$ \fi} 
\def \hhLsun {\ifmmode h^{-2}\,\rm L_{\odot} \else $h^{-2}\,\rm L_{\odot}$ \fi}

\def \LCDM {\ifmmode \Lambda{\rm CDM} \else $\Lambda{\rm CDM}$ \fi}
\def \sig8 {\ifmmode \sigma_8 \else $\sigma_8$ \fi} 
\def \OmegaM {\ifmmode \Omega_{\rm M} \else $\Omega_{\rm M}$ \fi} 
\def \OmegaL {\ifmmode \Omega_{\rm \Lambda} \else $\Omega_{\rm \Lambda}$\fi} 
\def \Deltavir {\ifmmode \Delta_{\rm vir} \else $\Delta_{\rm vir}$ \fi}

\def \rs {\ifmmode r_{\rm s} \else $r_{\rm s}$ \fi} 
\def \rrm2 {\ifmmode r_{-2} \else $r_{-2}$ \fi} 
\def \ccm2 {\ifmmode c_{-2} \else$c_{-2}$ \fi} 
\def \cvir {\ifmmode c_{\rm vir} \else $c_{\rm vir}$ \fi} 
\def \cbar {\ifmmode \overline{c} \else $\overline{c}$ \fi}

\def \R200 {\ifmmode R_{200} \else $R_{200}$ \fi} 
\def \Rvir {\ifmmode R_{\rm vir} \else $R_{\rm vir}$ \fi}

\def \v200 {\ifmmode V_{200} \else $V_{200}$ \fi} 
\def \Vvir {\ifmmode V_{\rm  vir} \else  $V_{\rm vir}$  \fi} 
\def  \Vhalo  {\ifmmode V_{\rm halo} \else $V_{\rm halo}$ \fi}

\def \M200 {\ifmmode M_{200} \else $M_{200}$ \fi} 
\def \Mvir {\ifmmode M_{\rm  vir} \else $M_{\rm  vir}$ \fi}  
\def \Mshell  {\ifmmode M_{\rm shell} \else $M_{\rm shell}$ \fi}

\def \Nvir {\ifmmode N_{\rm  vir} \else $N_{\rm  vir}$ \fi}  

\def \Jvir {\ifmmode J_{\rm vir} \else $J_{\rm vir}$ \fi} 
\def \Jshell {\ifmmode J_{\rm shell} \else $J_{\rm shell}$ \fi}

\def \Evir {\ifmmode E_{\rm vir} \else $E_{\rm vir}$ \fi} 

\def \lam {\ifmmode \lambda  \else $\lambda$ \fi} 
\def \lamp {\ifmmode \lambda^{\prime} \else $\lambda^{\prime}$  \fi} 
\def \lampc {\ifmmode \lambda^{\prime}_{\rm c} \else
  $\lambda^{\prime}_{\rm c}$  \fi} 
\def \lambar {\ifmmode \bar{\lambda}  \else  $\bar{\lambda}$  \fi}  
\def  \lampbar  {\ifmmode \bar{\lambda^{\prime}} \else
  $\bar{\lambda^{\prime}}$\fi} 
\def \siglam {\ifmmode \sigma_{\lambda} \else $\sigma_{\lambda}$ \fi} 
\def \siglamp {\ifmmode                \sigma_{\lambda^{\prime}} \else
$\sigma_{\lambda^{\prime}}$\fi}

\def \Rd {\ifmmode R_{\rm d} \else $R_{\rm d}$ \fi} 
\def \Rs {\ifmmode R_{\rm s} \else $R_{\rm s}$ \fi}  
\def \Rd {\ifmmode R_{\rm d} \else $R_{\rm d}$ \fi}  
\def \Rcool  {\ifmmode R_{\rm  cool}  \else $R_{\rm cool}$ \fi} 
\def \RIII {\ifmmode  3.2\Rs \else $3.2\Rs$ \fi} 
\def \RII {\ifmmode 2.2\Rs \else $2.2\Rs$  \fi} 
\def \Reff {\ifmmode R_{\rm eff} \else $R_{\rm  eff}$ \fi} 
\def  \rb {\ifmmode r_{\rm b}  \else $r_{\rm b}$ \fi}

\def  \Sigmacrit   {\ifmmode  \Sigma_{\rm  crit}   
\else  $\Sigma_{\rm crit}$\fi} 
\def \Sig0 {\ifmmode \Sigma_{0} \else $\Sigma_{0}$ \fi}

\def \muI {\ifmmode \mu_{0,I} \else $\mu_{0,I}$ \fi}

\def \mgal {\ifmmode m_{\rm gal} \else $m_{\rm gal}$ \fi} 
\def \md {\ifmmode m_{\rm d} \else $m_{\rm d}$ \fi} 
\def \ms {\ifmmode m_{\rm   s}   \else   $m_{\rm   s}$   \fi}   
\def   \mdbar   {\ifmmode {\overline{m}}_{\rm d} \else
  ${\overline{m}}_{\rm d}$ \fi} 
\def \msbar {\ifmmode  \bar{m}_{\rm  s}  \else  $\bar{m}_{\rm s}$
  \fi}  
\def  \Md {\ifmmode M_{\rm d}  \else $M_{\rm d}$ \fi} 
\def  \Ms {\ifmmode M_{\rm s} \else $M_{\rm  s}$ \fi} 
\def \Mb {\ifmmode  M_{\rm b} \else $M_{\rm b}$ \fi} 
\def \Mstar {\ifmmode  M_{\rm star} \else $M_{\rm star}$ \fi}
\def \Mdisc {\ifmmode M_{\rm disc} \else $M_{\rm disc}$ \fi}

\def \Jd {\ifmmode J_{\rm d} \else $J_{\rm d}$ \fi} 
\def \Jb {\ifmmode J_{\rm b} \else $J_{\rm b}$ \fi}  
\def \fb {\ifmmode  f_{\rm b} \else $f_{\rm b}$ \fi}

\def  \jd  {\ifmmode j_{\rm  d}  \else  $j_{\rm  d}$ \fi}  
\def  \jdmd {\ifmmode \frac{j_{\rm  d}}{m_{\rm d}} \else
  $\frac{j_{\rm d}}{m_{\rm d}}$ \fi} 
\def \fj {\ifmmode f_{\rm j} \else $f_{\rm j}$ \fi} 
\def \ft {\ifmmode f_{\rm t}  \else $f_{\rm t}$ \fi} 
\def  \fM {\ifmmode f_{\rm M} \else $f_{\rm M}$ \fi}

\def  \Vd {\ifmmode  V_{\rm  d}  \else $V_{\rm  d}$  \fi} 
\def  \Vcool {\ifmmode V_{\rm cool} \else $V_{\rm cool}$ \fi} 
\def \Vcirc {\ifmmode V_{\rm circ}  \else $V_{\rm circ}$  \fi} 
\def \VIII  {\ifmmode V_{3.2} \else $V_{3.2}$ \fi} 
\def  \VII {\ifmmode V_{2.2} \else $V_{2.2}$ \fi}
\def \Vobs {\ifmmode V_{\rm obs}  \else $V_{\rm obs}$ \fi} 
\def \Vdisc {\ifmmode V_{\rm disc} \else  $V_{\rm disc}$ \fi} 
\def \Vmax {\ifmmode V_{\rm  max} \else  $V_{\rm max}$  \fi} 
\def  \Vmaxobs{\ifmmode V_{\rm max}^{\rm obs}\else  $V_{\rm max}^{\rm
    obs}$\fi}  
\def \Vtot {\ifmmode V_{\rm tot} \else $V_{\rm tot}$  \fi} 
\def \Vrot {\ifmmode V_{\rm rot} \else  $V_{\rm rot}$  \fi} 
\def  \Vflat {\ifmmode  V_{\rm  flat} \else $V_{\rm flat}$ \fi}

\def \Ups {\ifmmode \Upsilon  \else $\Upsilon$ \fi} 
\def \YB {\ifmmode \Upsilon_B \else $\Upsilon_B$ \fi} 
\def \YI {\ifmmode  \Upsilon_I  \else $\Upsilon_I$ \fi} 
\def \DeltaIMF {\ifmmode \Delta_{\rm IMF} \else $\Delta_{\rm IMF}$ \fi}

\def\LCDM{$\Lambda$CDM }

\def\c200{$c_{200}$}

\title[Milky Way Satellites in a \LCDM Universe] {Luminosity function and 
radial distribution of Milky Way Satellites in a \LCDM Universe}

\author[A.V.  Macci\`o et al.]  {Andrea
  V. Macci\`o$^{1}$\thanks{maccio@mpia.de}, Xi Kang$^{1}$, Fabio
  Fontanot$^1$, Rachel S. Somerville$^{1,2}$ \newauthor{Sergey
    Koposov$^{1,3,4}$, Pierluigi Monaco$^{5,6}$}\\ $^1$
  Max-Planck-Institut f\"ur Astronomie, K\"onigstuhl 17, 69117
  Heidelberg, Germany \\ $^2$ Space Telescope Science Institute, 3700
  San Martin Drive, Baltimore, MD 21218, USA \\ $^3$ Institute of
  Astronomy, University of Cambridge, Madingley Road, Cambridge,
  UK\\ $^4$ Sternberg Astronomical Institute, Universitetskiy pr. 13,
  119992 Moscow, Russia \\ $^5$ Dipartimento di Astronomia,
  Universit\`a di Trieste, via Tiepolo 11, 34131 Trieste, Italy
  \\ $^6$ INAF-Osservatorio Astronomico, Via Tiepolo 11, I-34131
  Trieste, Italy \\ }

\begin{document}
             
\date{submitted to MNRAS}
             
%\pagerange{\pageref{firstpage}--\pageref{lastpage}}\pubyear{2006}

\maketitle           

\label{firstpage}
             
\begin{abstract}

We study the luminosity function and the radial distribution
of satellite galaxies within Milky
Way sized haloes as predicted in Cold Dark Matter based models of
galaxy formation, making use of numerical N-body techniques 
as well as three different semi-analytic
model (SAMs) galaxy formation codes. We extract merger trees from very
high-resolution dissipationless simulations of four
Galaxy-sized DM haloes, and use these as common input for the
semi-analytic models. We present a detailed comparison of our
predictions with the observational data recently obtained on the Milky
Way satellite luminosity function (LF).  We find that semi-analytic
models with rather standard astrophysical ingredients are able to
reproduce the observed luminosity function over six orders of
magnitude in luminosity, down to magnitudes as faint as $M_V=-2$.  We
also perform a comparison with the actual observed number of
satellites as a function of luminosity, by applying the selection
criteria of the SDSS survey to our simulations instead of correcting
the observations for incompleteness.  Using this approach we again
find good agreement for both the luminosity and radial distributions
of MW satellites. We investigate which physical processes in our
models are responsible for shaping the predicted satellite LF, and
find that tidal destruction, suppression of gas infall by a
photo-ionizing background, and supernova feedback all make important
contributions.  We conclude that the number and luminosity of Milky Way
satellites can be naturally accounted for within 
the ($\Lambda$)Cold Dark Matter paradigm, and this should 
no longer be considered a problem.
\end{abstract}

\begin{keywords}
galaxies: haloes -- cosmology:theory, dark matter, gravitation --
methods: numerical, N-body simulation
\end{keywords}

\setcounter{footnote}{1}

%%%%%%%%%%%%%%%%%%%%%%%%%%%%%%%%%%%%%%%%%%%%%%%%%%%%%%%%%%%%%%%%%%%%%%
%% SECTION 1: INTRODUCTION
%%%%%%%%%%%%%%%%%%%%%%%%%%%%%%%%%%%%%%%%%%%%%%%%%%%%%%%%%%%%%%%%%%%%%%
\section{Introduction}
\label{sec:intro}

The Milky Way environment provides an excellent laboratory for
astrophysics. It has been used extensively in the past decades to test
theoretical models of galaxy formation.  In particular, the number
density of satellites around our Galaxy has long been considered one
of the major problems for the otherwise quite successful $\Lambda$CDM
paradigm.

About a decade ago, N-body simulations attained sufficient dynamic
range to reveal that in CDM models, all haloes should contain a large
number of embedded subhaloes that survive the collapse and
virialization of the parent structure (Klypin \etal 1999; Moore \etal
1999 and more recently Diemand \etal 2007). Although the predicted
number of substructures was in reasonable agreement with observed
luminosity functions in cluster sized haloes, in Milky Way sized
haloes the number of predicted sub-haloes exceeded the number of
observed satellites by at least an order of magnitude: the known
satellite population at that time consisted of about 40 satellites with
$V_c \gta 20$ km/s in the Local Group (e.g. Mateo 1998), while the
simulations predicted about 300 sub-haloes with $V_c \gta 20$ km/s
(Klypin \etal 1999; Moore \etal 1999).

Several astrophysical solutions to this problem have been proposed.
Many authors have pointed out that accretion of gas into low-mass
haloes and subsequent star formation is inefficient in the presence of
a strong photoionizing background, as this background radiation raises
the entropy of the gas, preventing it from accreting onto small dark
matter haloes and lengthening the cooling time of that gas which has
accreted (e.g. Babul \& Rees 1992; Efstathiou 1992; Thoul \& Weinberg
1996; Quinn, Katz, \& Efstathiou 1996). Several studies showed
quantitatively that this suppression of gas infall by cosmic
reionization could plausibly reconcile the observed and predicted
numbers of ``classical'' Local Group satellites
(Bullock \etal 2000, Somerville 2002, Ricotti, Gnedin \& Shull 2002,
Benson \etal 2002, Read \etal 2006). It was also pointed out that tidal stripping and
heating as satellites orbited in the potential of the larger galaxy
could cause dramatic mass loss, even decreasing the circular velocity
in the inner parts of the sub-halo (Kravtsov, Gnedin \& Klypin 2004;
Taylor \& Babul 2004; Zentner \etal 2005). Thus many Local Group satellites may inhabit dark matter
(sub-)haloes that are much less massive than they were at the time that
they were accreted by their host halo.

In recent years the Sloan Digital Sky Survey (SDSS: Adelman-McCarthy
\etal 2008) has changed our view of the Milky Way and its
environment. The SDSS has made it possible to carry out a systematic
survey for satellite galaxies, which are detectable through their
resolved stellar populations down to extremely low surface
brightness. As a result the number of known dwarf spheroidals has
doubled in the recent past (e.g Willman \etal 2005, Zucker \etal 2006;
Belokurov \etal 2007; Irwin \etal 2007, Gilmore \etal 2007).
Spectroscopic surveys subsequently measured the velocity dispersions
of these systems, and confirmed their galactic nature (Martin \etal
2007, Simon \& Geha 2007).  This recently discovered population of
ultra faint satellites has posed new challenges for models of galaxy
formation and opened the possibility to test the \LCDM paradigm at
very small mass scales (e.g. Strigari \etal 2008, Macci\`o \etal
2009).

These new observations have made it possible to probe the faint end of
the luminosity function of Milky Way satellites, down to luminosities
as faint as 100 $L_{\odot}$.  Moreover the homogeneous sky coverage of
the SDSS enables a robust determination of the detection limits for
faint satellites.  Koposov \etal (2008) provided the first
determination of the volume corrected Milky Way satellite luminosity
function down to these extremely faint limits, by assuming various simple
radial distribution functions for the satellite population and
applying the SDSS detection limits. 

In light of the discovery of the new ultra-faint dwarf population and
the improvements in the numerical modelling of galaxy formation, it is
now timely to revisit the issue of whether the basic properties of
satellite galaxies around the Milky Way, such as their number density,
radial distribution, and mass-to-light ratios, can be reproduced
within current cosmological $\Lambda$CDM-based models. It is also
interesting to ask what physical processes might plausibly give rise
to this population of extremely low-luminosity galaxies.

In this paper we combine merger trees extracted from very high
resolution N-body simulations with three different semi-analytic model
(SAM) codes. These merger trees describe the hierarchical assembly of
a Milky Way-like halo, while the SAMs are used to predict the
relationship between the dark matter (sub)haloes and observable galaxy
properties, allowing us to make a direct and detailed comparison with
observational data. 

The goal of this work is not only to test whether the observed
properties of Milky Way satellites, including the recently discovered
faint population, can be reproduced within the \LCDM model, but
also to understand how and when this extreme population formed. We aim
to understand how various mechanisms (such as SN feedback, cosmic
photo-ionization, and tidal stripping) may shape the luminosities 
of galaxies populating low mass dark matter
substructures orbiting around Milky Way-like galaxies.

The remainder of this paper is organized as follows. In Section
\ref{sec:sims} we describe the numerical simulations.  In Section
\ref{sec:sam} we briefly summarize the SAMs used in our study,
highlighting the differences among the models.  Section \ref{sec:obs}
contains a detailed description of the observational data used in this
work.  In Section \ref{sec:res} we compare the luminosity function,
and radial distribution of simulated satellites
with observational data. Finally in Section \ref{sec:concl} we present
our main conclusions.

%%%%%%%%%%%%%%%%%%%%%%%%%%%%%%%%%%%%%%%%%%%%%%%%%%%%%%%%%%%%%%%%%%%%%%
%% SECTION 2: N-BODY SIMULATIONS
%%%%%%%%%%%%%%%%%%%%%%%%%%%%%%%%%%%%%%%%%%%%%%%%%%%%%%%%%%%%%%%%%%%%%%
\section{Simulations} 
\label{sec:sims}

The N-body simulations of this study were obtained using {\sc pkdgrav},
a treecode written by Joachim Stadel and Thomas Quinn (Stadel 2001).
The initial conditions are generated with the {\sc grafic2} package
(Bertschinger 2001).  The starting redshift $z_i$ is set to the time
when the standard deviation of the smallest density fluctuations
resolved within the simulation box reaches $0.2$ (the smallest scale
resolved within the initial conditions is defined as twice the
intra-particle distance).  The cosmological parameters are chosen to
be: $\Omega_{\Lambda}$=0.732, $\Omega_m$=0.268, $\Omega_b$=0.044,
$h=0.71$ and $\sigma_8=0.9$, and are in reasonable agreement with the
recent WMAP mission results (Komatsu \etal 2009).

We selected four candidate haloes with a mass similar to the mass of
our Galaxy ($M \sim 10^{12} \Msun$) from an existing low resolution
dark matter simulation (300$^3$ particles within 90 Mpc) and
re-simulated them at higher resolution.  Our high resolution haloes
all have a quiet merging history with no major merger after $z=2$, and
thus are likely to host a disk galaxy at the present time (with the
exception of G3, which we discuss further below).
The standard high resolution runs are 12$^3$ times more resolved in
mass than the initial simulation: the dark matter particle mass is
$m_{d} = 4.16 \times 10^5 \hMsun$, where each dark matter particle has
a spline gravitational softening of 355 $h^{-1}$ pc. Some of the main
properties of the re-simulated haloes are listed in Table
\ref{table:haloes}.  One of the haloes, namely G3, has a mass greater
than the expected mass of the MW and has experienced a major merger at
z=1.5 so it is likely to host an elliptical galaxy.  In order to check
possible resolution effects (especially in the construction of the
merger tree) we re-simulated one of the haloes (namely G1) with higher
resolution (27$^3$ times with respect to the low resolution), with
more than 32 million particles within the virial radius (G1$_{\rm HR}$ in
Table \ref{table:haloes}), reaching a dark matter particle mass of $m
= 3.65 \times 10^4 \hMsun$.

\begin{table}
 \centering
 \begin{minipage}{140mm}
  \caption{Dark Matter Halo parameters}
  \begin{tabular}{lccccc}
\hline  Halo &  Mass  & N$_{\rm part}$  & $R_{\rm vir}$ & $V_{\rm circ}$  \\
       &($10^{12}\hMsun)$& $(10^6)$       &  (kpc/h)          & (km/s)   \\
\hline 
G0   & 0.88  & 2.12   & 197 & 178  \\
G1   & 1.22  & 2.93   & 219 & 188  \\
G2   & 1.30  & 3.12   & 250 & 203  \\
G3   & 2.63  & 5.64   & 268 & 236  \\
G1$_{\rm HR}$ & 1.15  & 31.5   & 211  & 184 \\
\hline 
\label{table:haloes}
\end{tabular}
\end{minipage}
\end{table}

For the purpose of constructing accurate merger trees for each
simulated halo, we analyse 53 output times between $z=20$ and $z=0$.
For each snapshot, we look for all the virialized isolated haloes within the
high resolution region using a Spherical Overdensity (SO) algorithm.
We use a time varying virial density contrast determined
using the fitting formula presented in Mainini \etal (2003).
We include in the halo catalogue all the haloes with more than 100 particles 
(see Macci\`o \etal 2007, 2008 for further details on our halo finding
algorithm). Our procedure to construct merger trees is described in detail 
in Macci\`o, Kang \& Moore (2009).
We used all particles within 1.5 times the virial radius of a
given ``root'' halo at $z=0$ and then track them back to the previous
output time. We then make a list of all haloes at that earlier output
time containing marked particles, recording the number of marked
particles contained in each one. We use the two criteria suggested 
in Wechsler \etal (2002) for halo 1 at one output time to be 
labeled a ``progenitor'' of halo 2 at the subsequent output time.
In our language, halo 2 will then be labeled
as a ``descendant'' of halo 1 if i) more than 50\% of the particles in
halo 1 end up in halo 2 or if ii) more than 75\% of halo 1 particles
that end up in any halo at time step 2 end up in halo 2 (this second
criterion is mainly relevant during major mergers).  Thus a halo can
have only one descendant but there is no limit to the number of
progenitors. On average there are ~20,000 progenitors for haloes
G0-G3, while the number of progenitors for the G1$_{\rm HR}$ run is close
to 100,000.

In order to identify subhalos in our simulation we have run the
MPI+OpenMP hybrid halo finder {\sc AHF} ({\sc AMIGA} halo finder, to
be downloaded freely from http://www.popia.ft.uam.es/AMIGA) described
in detail in Knollmann \& Knebe (2009). {\sc AHF} locates local
overdensities in an adaptively smoothed density field as prospective
halo centres.  The local potential minima are computed for each of
these density peaks and the gravitationally bound particles are
determined. Only peaks with at least 50 bound particles are considered
as haloes and retained for further analysis.  As subhaloes are
embedded within their respective host halo, their own density profile
usually shows a characteristic upturn at a radius $r_t \lta r_{\rm
  vir}$, where $r_{\rm vir}$ would be their actual (virial) radius if
they were found in isolation.  We use this ``truncation radius'' $r_t$
as the outer edge of the subhalos and hence (sub-)halo properties
(i.e. mass) are calculated using the gravitationally bound particles
inside $r_t$.

 %%%%%%%%%%%%%%%%%%%%%%%%%%%%%%%%%%%%%%%%%%%%%%%%%%%%%%%%%%%%%%%%%%%%%%%%
%% SECTION 3: SAMs
%%%%%%%%%%%%%%%%%%%%%%%%%%%%%%%%%%%%%%%%%%%%%%%%%%%%%%%%%%%%%%%%%%%%%%%%

\section{Semi-Analytic Models}
\label{sec:sam}

We make use of three different semi-analytic model (SAM) codes in
order to predict the observable properties of galaxies that inhabit
the dark matter haloes and sub-haloes identified in the $N$-body
simulations described above (see Baugh 2006 for a recent review on the
semi-analytic approach). We will consider predictions from the most
recent implementations of three different SAMs, developed
independently by different groups: (i) the Kang \etal (2005) model
that has been recently updated in Kang (2008, K08 hereafter); (ii) the
fiducial model of Somerville \etal (2008, S08 hereafter), which builds
on the original formulation presented in Somerville \& Primack (1999)
and Somerville \etal (2001); (iii) {\sc morgana}, first presented in
Monaco, Fontanot \& Taffoni (2007) and then updated in Lo Faro \etal
(2009). Since all SAMs assume that DM haloes are the sites where
galaxy formation takes place and they need a proper description of
their assembly history, we will use the four merger trees extracted
from N-body simulations of the G0-G3 haloes (see section
\ref{sec:sims}) as a common input. In order to increase the
statistical robustness of our results, in sec.~\ref{ssec:lf} we also
consider a larger set of realizations of merger trees obtained using
the extended Press-Schechter (EPS) formalism (e.g. Somerville \&
Kolatt 1999, Parkinson \etal 2008) for K08 and S08 and the lagrangian
code {\sc pinocchio} (Monaco \etal 2002) for {\sc morgana}.

All the SAMs considered in this work parametrize in different ways the
main physical processes acting on the baryonic component, such as
atomic cooling, cosmic reionization, star formation, supernovae
feedback, metal production and dust attenuation. For sake of
simplicity, we will discuss here only those processes relevant in
shaping the LF of MW satellites.  We refer the reader to the original
works for a more detailed discussion on the modeling of physical
processes (see also Fontanot et al. 2009, for a comparison between
different SAMs).

Although our simulations resolve subhaloes \footnote{From this point
  on, we refer to the DM haloes living within the virial radius of
  larger haloes as ``substructure'' or ``subhaloes'', while we refer
  to the all the galaxies except the central galaxy of the larger halo
  as ``satellites''.}
we do not record the fate of subhaloes in our merger trees or make use
of this information in the SAMs. When a subhalo is accreted, its
position is initially either set equal to the virial radius of the
parent halo at that time (K08 and S08), or extracted from a suitable
distribution of radial distances (\morgana). Moreover the orbital
parameters (velocity and orbit eccentricity) for each infalling
satellite are randomly selected from a distribution motivated by the
statistics of satellite orbits in cosmological simulations. The
dynamical evolution (and so the survival probability) of each subhalo
is then computed by estimating the time required for the subhalo to
lose all of its orbital energy due to dynamical friction against the
background DM potential (using updated variants of the classical
Chandrasekhar formula).

Each of the models that we have considered applies a different set of
criteria to determine when satellites are destroyed by tidal
stripping. In the K08 models, a subhalo is considered to be tidally
destroyed if it either loses more than 98\% of its mass
(e.g. Penarrubia \etal 2008) or if its mass falls below $6.5\times
10^6 \Msun$, which is the minimum mass observed for Milky Way
satellites (Strigari \etal 2008).
In the S08 model, satellites are considered to be tidally destroyed
when their stripped mass drops below the mass contained within a fixed
fraction of the halo's original NFW (Navarro, Frenk and White 1997)
scale radius $r_s$ (following Zentner \& Bullock 2003 and Taylor \&
Babul 2004).
In \morgana, the tidal radius is computed at the first periastron of
the satellite orbit by computing the radius at which the density of
the unperturbed satellite is equal to the density of the main DM halo
at the periastron. All the mass (whether dark, stellar, or gaseous)
external to the tidal radius (i.e. at a lower density) is then
considered unbound. The \morgana~ estimates of the radii of the bulge
and disk components, plus assumed density profiles for the stars and
gas, are used to estimate the fraction of the baryonic mass that lies
outside the tidal radius.

In all three models the effect of reionization is expressed in terms
of a ``filtering mass'' (e.g. Gnedin 2000). This filtering mass
corresponds to the mass at which haloes will only be able to accrete
half of the universal baryonic content.  The fraction of baryons that
can be accreted as hot gas is parameterized using the following
expression (Gnedin 200):
\begin{equation}
f_{\rm b, acc}(z, M_{\rm vir}) = \frac{f_{b}}{[1 + 0.26 M_F(z)/M_{\rm
      vir}]^3}\, ,
\label{eqn:fbar}
\end{equation}
where $f_{b}$ is the universal baryon fraction and $\Mvir$ is the halo
virial mass. The filtering mass as a function of redshift $M_F(z)$
depends on the reionization history of the Universe, and is
parameterized using the fitting formulae provided by Kravtsov \etal
(2004, but see section \ref{ssec:reion}).

Massive stars and supernovae may impart thermal and kinetic energy to
the cold interstellar medium: in the K08 and S08 models, the rate of
reheating of cold gas due to supernova feedback is given by an
expression of the form:
\begin{equation}
\dot{m}_{\rm rh} = \epsilon^{SN}_0 \left( \frac{V_{\rm disk}}{V_0}
\right)^{\alpha_{\rm rh}} \, \dot{m}_*
\label{eqn:snfb}
\end{equation}
where $\epsilon^{SN}_0$ and $\alpha_{\rm rh}$ are free parameters and
$\dot{m}_*$ is the star formation rate. K08 and S08 adopt similar
values of $\epsilon^{SN}_0$ and $\alpha_{\rm rh} \sim -2$, chosen to
reproduce the faint end slope of the observed $z=0$ global galaxy
luminosity function or the low-mass end of the stellar mass function.
In contrast, the \morgana~ model adopts a recipe based on the notion
of a self-regulated feedback loop between star formation and
supernovae (Monaco 2004), which roughly corresponds to
$\epsilon_0^{SN}=1$ and $\alpha_{\rm rh}=0$ in terms of
eqn.~\ref{eqn:snfb}. However, we find that in order to reproduce the
MW satellite LF with \morgana~ we need to introduce a strong
dependence of the mass loading factor $\eta \equiv \dot{m}_{\rm
  rh}/\dot{m}_*$ on the galaxy circular velocity\footnote{In order not
  to spoil the agreement of this model for $M_\star > 10^9 M_\odot$
  galaxies, we retain the original recipe $V_c > 100$ km/s DM halos
  and adopt Eqn.~\ref{eqn:snfb} in galaxies with $V_c < 100$ km/s.}
($\alpha_{\rm rh}=-4$). This suggests that the phenomenological
scaling in eqn.~\ref{eqn:snfb} is a key to the success of galaxy
formation models in the $\Lambda$CDM context in predicting the MW
satellite LF.

It is worth noting that the ingredients of these SAMs have all been
developed with much larger galaxies in mind, and the models have
previously been calibrated mainly against observations of relatively
luminous galaxies ($M_* \gta 10^9 M_\odot$ or $M_V \lta -16$). It is
quite unclear whether the standard semi-analytic empirical recipes for
e.g. star formation or supernova feedback should apply in galaxies as
tiny as the ultra-faint Milky Way dwarfs, which may form out of just a
few molecular clouds. Therefore, it is quite an interesting experiment
to see how well these models perform when extended to these very
different mass scales.

%%%%%%%%%%%%%%%%%%%%%%%%%%%%%%%%%%%%%%%%%%%%%%%%%%%%%%%%%%%%%%%%%%%%%%%%
%% SECTION 3b: Observational data 
%%%%%%%%%%%%%%%%%%%%%%%%%%%%%%%%%%%%%%%%%%%%%%%%%%%%%%%%%%%%%%%%%%%%%%%%
\section{Observational Data}
\label{sec:obs}

We test our MW models against observations by focusing on two key
aspects of MW satellite galaxy properties: their luminosity
and radial distributions.

For the luminosity function we use the results of Koposov \etal (2008;
SK08 hereafter). SK08 recently presented a quantitative search
methodology for Milky Way satellites in the SDSS DR5 data and used
this method to compute detection efficiency maps, which ultimately
allowed the construction of the first completeness-corrected satellite
galaxy luminosity function (see also Walsh, Willman \& Jergen 2009).
SK08 measured the luminosity function (LF) down to $M_V=-2$, and found
that it can be described by a single power law of the form $dN/dM_v=10
\times 10^{0.1 (M_V+5)}$.  At the very faint end ($M_V >-5$), in order
to compute the completeness correction, a radial satellite
distribution around the host must be assumed; in all the luminosity
function plots presented in this paper the upper data points (always
shown as open circles with no error bars) are obtained assuming an
isothermal density distribution while the lower points (solid circles
with error bars) are obtained assuming an NFW (Navarro, Frenk and
White 1997) distribution.

We also use a reverse approach to addressing the completeness issue by
performing the comparison in ``observational space''. Instead of
assuming a radial distribution for the observed galaxies, we apply the
detection criteria of the SDSS to our simulations (see section
\ref{ssec:SDSS} for more details) and compare directly with the raw
data from the SDSS. For this comparison we construct a ``hybrid'' data
set. For satellites brighter than $M_V=-9$, in order to increase the
number statistics, we gather together satellites from the Milky Way
and the Andromeda galaxy (data from Mateo 1998 and Metz \etal 2007),
and we assign a weight of $w=0.5$ to each satellite, assuming that
current surveys are complete down to this limit.  For fainter
satellites we collect data from Martin, de Jong \& Rix 2008 (MdJR08
hereafter) and, in order to account for the fact that the SDSS
surveyed only one-fifth of the sky, we set $w=5$ for these faint
galaxies. The adopted $M_V$ threshold for splitting the observational
sample is justified by the low luminosity of all newly discovered
satellites both around the Milky Way and the Andromeda galaxy
(McConnachie \etal 2008).

In addition, using the same data set described above, we compute the
cumulative radial distribution of satellites (i.e. the number of
satellites within a given distance from the Sun).  Distances for
bright satellites are taken from Metz \etal (2007), while we use
results compiled in MdJR08 for faint galaxies ($M_V>-9$).  In
computing the radial distribution we assign to each galaxy the same
weight adopted for the LF.

%%%%%%%%%%%%%%%%%%%%%%%%%%%%%%%%%%%%%%%%%%%%%%%%%%%%%%%%%%%%%%%%%%%%%%%%
%% SECTION 4: Results SAMs
%%%%%%%%%%%%%%%%%%%%%%%%%%%%%%%%%%%%%%%%%%%%%%%%%%%%%%%%%%%%%%%%%%%%%%%%
\section{Results}
\label{sec:res}

In this section we present results for both the SAMs and numerical
(dissipational and dissipationless) simulations and compare them with
the observational data set described in Section \ref{sec:obs}.  First
we compare the dynamical evolution of satellites in SAMs and in N-body
simulations, and then we present results for the Luminosity Function
(LF) of simulated satellites and analyze the importance of different
physical processes (e.g. reionization, stellar stripping and supernova
feedback) in shaping the LF.  We then present results for 
the radial distribution and  compare them with observations.

\subsection{Dynamical evolution of sub-haloes in SAMs and N-body simulations}
\label{ssec:cfrnbody}

\begin{figure}
\psfig{figure=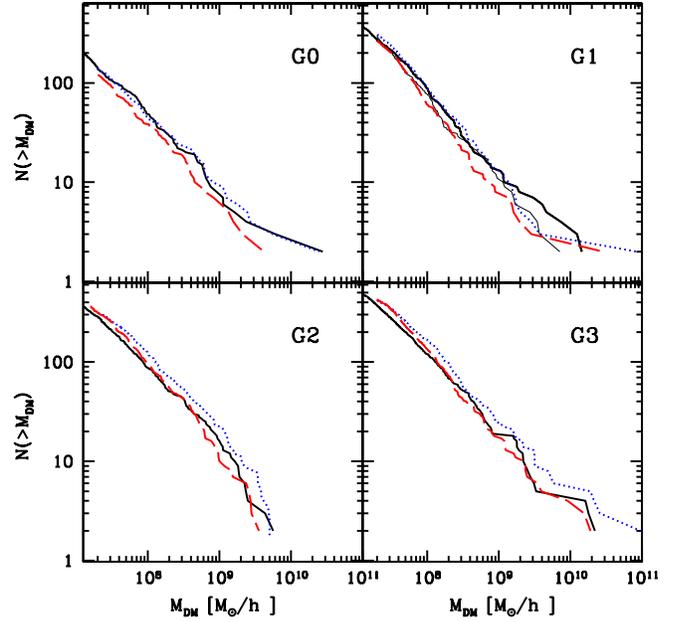,width=0.5\textwidth}
\caption{\scriptsize Comparison between the subhalo mass function at
  $z=0$ (within \Rvir) in the N-body simulations (solid black line)
  and the predictions from the semi-analytic models of K08 (red dashed
  line) and S08 (blue dotted line).  Each panel shows results for a
  different dark matter halo. The (black) thin solid line in the G1 panel
  shows the results of the K08 SAM when applied to the G1$_{\rm HR}$
  halo.}
\label{fig:mfDM}
\end{figure}

In the SAMs investigated here, we chose to make use of N-body based
merger trees for ``isolated'' (or distinct) haloes only, and to model
the dynamical evolution of satellites semi-analytically (see
Section~\ref{sec:sam}). Thus, there will not be a one-to-one
correspondence between the masses or positions of sub-haloes at $z=0$
in the SAMs and in the actual N-body simulations. In this section we
check that the statistical distributions of subhalo masses and radii
predicted by the semi-analytic models are in agreement with those of
subhaloes identified in the N-body simulations. This comparison is
only possible for the K08 and S08 models, since the \morgana~ model
does not explicitly follow the dynamics of dark matter substructures.
Figure \ref{fig:mfDM} shows the cumulative subhalo mass function from
N-body and SAMs for our four Merger tree G0-G3. The G1$_{\rm HR}$ halo
is shown for the K08 model only. The SAM results are obtained by
averaging over ten different realizations of the random orbit
selection process and are truncated at the N-body mass resolution
limit.  A simple way to quantify the agreement between two distributions is to 
perform a Kolmogorov-Smirnov (KS) test (Press et al. 1992). 
In the following we will quote as KS results,  the probability that two 
distributions are drawn from the same parent population. 
K08 and S08 SAMs give an average KS probability of 88\% and 90\%  
respectively, when compared to the subhalo mass function 
from the N-body simulations, showing thus good agreement with the 
numerical results.

%-----------------------------------------------------------------------------------------------
%-----------------------------------------------------------------------------------------------

\begin{figure}
\psfig{figure=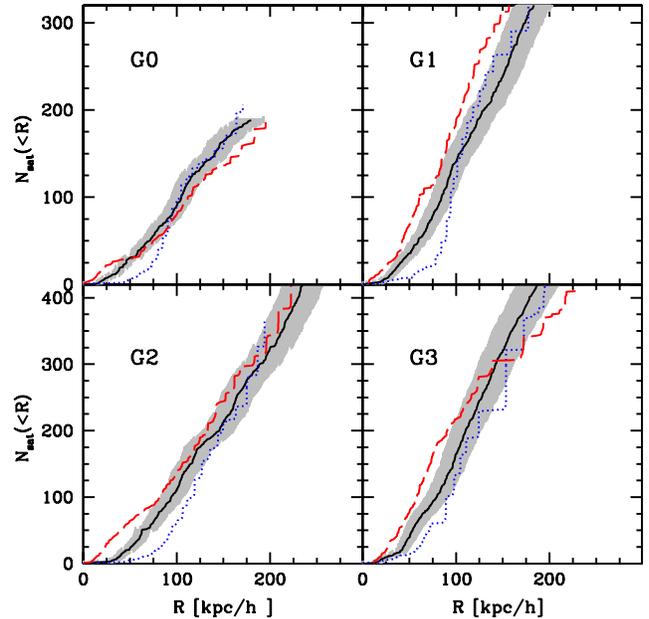,width=0.5\textwidth}
\caption{\scriptsize The cumulative number density of satellites as a
  function of radius from the center of the parent galaxy.  The
  results from the N-body simulations are shown by the solid (black)
  line, the shaded (grey) area shows the 1$\sigma$ scatter over five
  different realizations of the satellite orbit distribution.
 Results from the K08 and S08 SAMs are shown by the (red) dashed line
 and (blue) dotted line respectively.  Each panel shows results for a
 different dark matter halo.}
\label{fig:radDM}
\end{figure}

The radial distribution of satellites is a key piece of information
for deciding whether a satellite will be detectable in the SDSS
survey. It is therefore important to also check that the SAMs
correctly predict the radial distribution of subhaloes within the
parent halo.  Figure \ref{fig:radDM} shows the radial number density
of subhaloes (without including the central galaxy); 
while on average there is good agreement between SAMs and N-body results 
for both models (KS test results: 87\%  and 88\% S08 and K08 respectively),
there is a systematic off-set between K08 and S08, especially at distances $<100$
kpc/h. The reason for this off-set resides in the 
different fitting formulas for tidal destruction and dynamical time 
implemented in the two models, but is nevertheless 
too small to affect our analysis.

%---------------------------------------------------------------------------
%---------------------------------------------------------------------------------------

In the semi-analytic models used here, the initial orbit eccentricity
of an infalling subhalo is randomly selected from a distribution
motivated by the results from cosmological N-body simulations.  We
checked that the scatter due to different realizations of the orbit
distribution is fairly small, and mostly effects the less numerous,
massive satellites which are not the focus of our study.

\subsection{Satellite Luminosity Function}
\label{ssec:lf}

\begin{figure*}
\psfig{figure=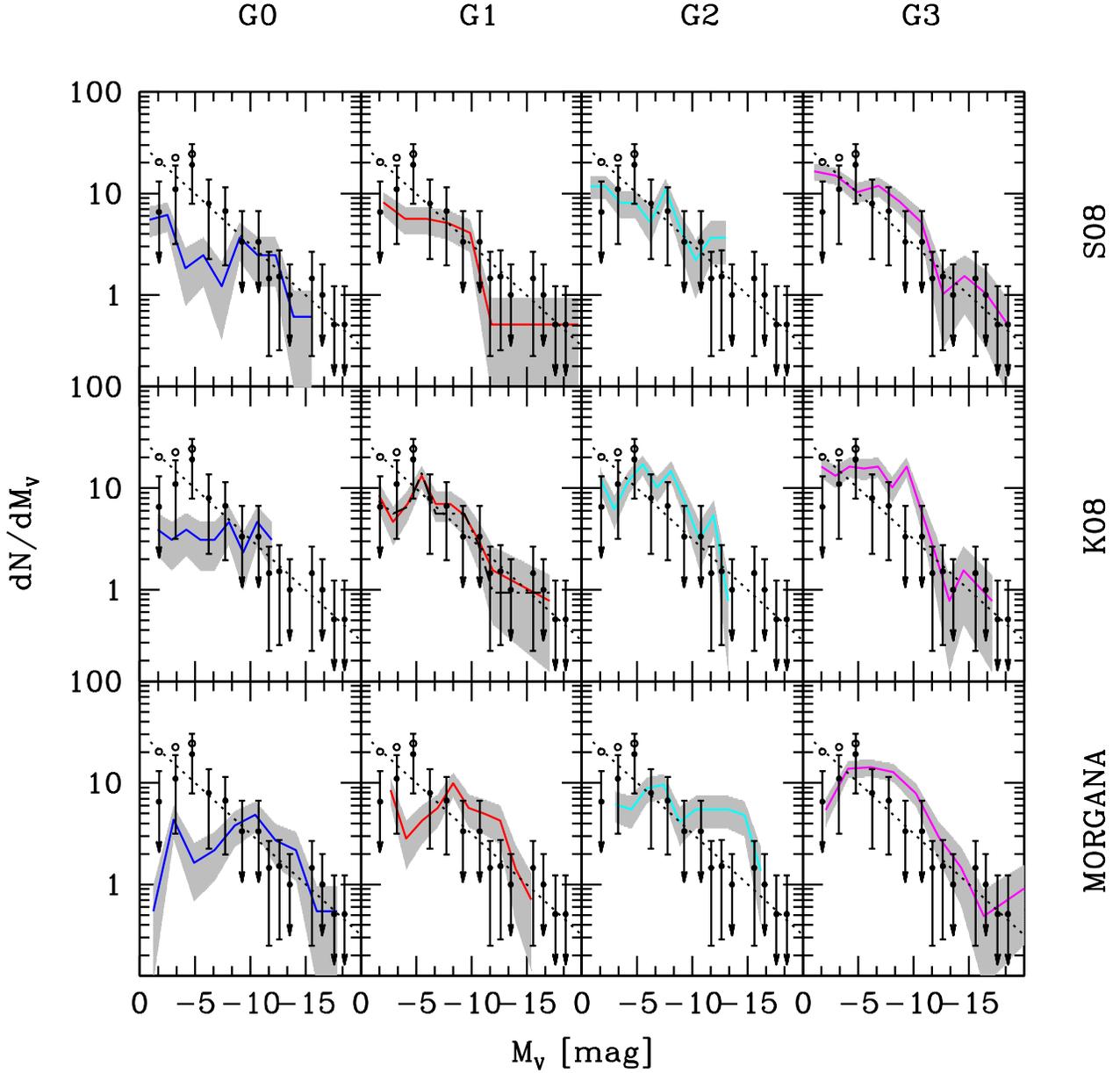,width=500pt}
\caption{\scriptsize The Milky Way satellite luminosity function
  predicted by our three semi-analytic models (S08, K08 and \morgana~
  from top to bottom) using the G0-3 halo merger trees (from left to
  right), with an assumed reionization redshift $z_r=7.5$.  The median
  of the satellite distribution is shown by the solid line, while the
  shaded area represents the $1\sigma$ Poisson scatter around the
  mean.  Observational data are taken from SK08 under the assumption
  of two different radial distributions of satellites, NFW-like (solid
  circles with error bars) and isothermal (open circles). The arrows
  on error bars indicate that there is only one galaxy in that
  particular bin, and so the Poisson error is formally 100\%. The
  dotted line shows a single power law fit to the data:
  $dN/dM_v=10\times 10^{0.1(M_V+5)}$. In the G1 panel of the K08 model
  results for the G1$_{\rm HR}$ run are also shown as a dashed (black)
  line.}
\label{fig:MvAll}
\end{figure*}

We now compute satellite LFs using the four merger trees obtained from
the N-body simulations as common input for our semi-analytic models;
this allows us to isolate the impact of the different physical
ingredients in the SAMs because the DM halo merger histories are
exactly the same in all three models. Figure~\ref{fig:MvAll} shows the
predictions from our three SAMs, adopting $z_r=7.5$ as the redshift of
reionization. We plot all satellites within $R=280$ kpc in order to be
consistent with the SK08 data set.  
Solid lines show the mean of satellite distribution and the (grey)
shaded area shows the 1$\sigma$ Poisson scatter around that mean.  

All of the models predict that the number of satellites brighter than
$M_V=-3$ is of the order of $\sim 100$ and could easily be several
hundreds (e.g G3).
This is in agreement with recent estimates of the number of observed
satellites obtained by several different approaches (e.g. Tollerud
\etal 2008; Madau \etal 2008; Koposov \etal 2009).

%----------------------------------------------------------------------------
%-----------------------------------------------------------------------

All the models show that the total number of satellite galaxies
depends on the host halo mass.  The LF of the G0 halo (the least
massive one, see Table \ref{table:haloes}) is almost flat and has a
lack of satellites at the faint end compared to the MW data. On the
other hand, in the most massive halo (G3), the SAMs predict more
satellite galaxies at all luminosities than are observed in the
MW. This trend between halo mass and the LF does not depend on the SAM
used to populate dark matter substructures and indicates that the
total mass of the dark matter halo has a quite strong influence in
determining the normalization (and shape) of the satellite LF.

All three semi-analytic models considered in this work are able to do
a reasonably good job of reproducing the observational data.  The K08
and S08 models (upper and middle panels) in figure \ref{fig:MvAll}
quite successfully reproduce the observed LF for satellite galaxies
over the entire luminosity range $-2 \geq M_V \geq -16$.  The \morgana~
model (lower panels in figure \ref{fig:MvAll}), tends to predict
slightly more satellites at intermediate luminosity ($-10 > M_V >
-12$) but it is nonetheless in good agreement with the
observations especially if an NFW distribution is assumed for observed
satellites. In the G1 panel for the K08 model we also show the LF
obtained using the G1$_{\rm HR}$ merger tree (dotted black line). We see
that the higher resolution merger tree produces almost
indistinguishable results.

\begin{figure}
\psfig{figure=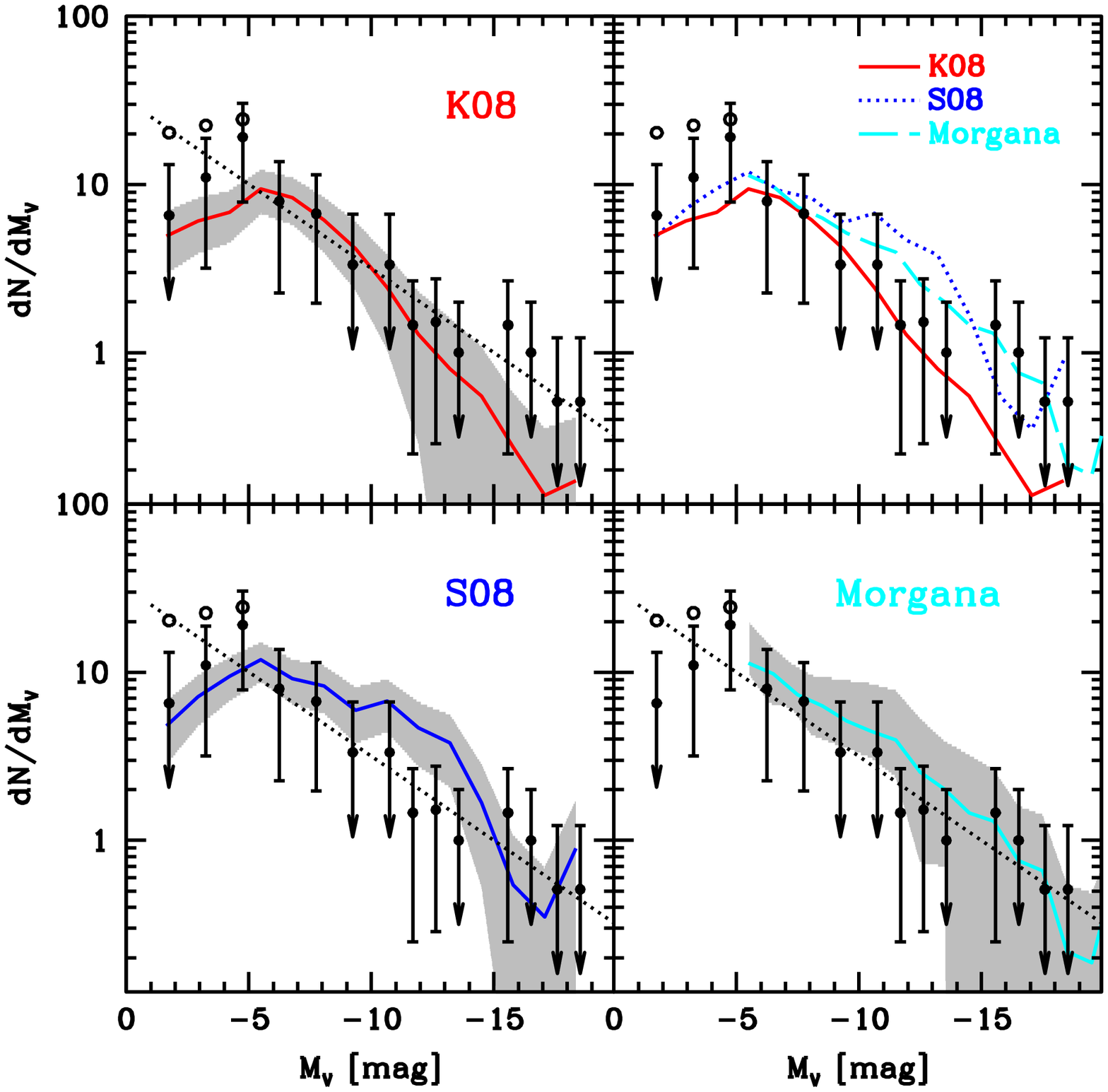,width=0.5\textwidth}
\caption{\scriptsize Satellite Luminosity function predicted by SAMs
  using semi-analytic merger trees (EPS and {\sc pinocchio}, see text
  for details).  Symbols have the same meaning as in figure
  \ref{fig:MvAll}.  The upper right panel shows the average satellite
  LF for the three semi-analytic models together.  \morgana~ results
  are shown only down to $M_V=-5$, due to a resolution limitation in
  the {\sc pinocchio} code; see text for more details.  }
\label{fig:AveMv}
\end{figure}

In some cases, certain models and certain haloes show a dearth or even
absence of the most luminous satellites (e.g. S08 and K08 G0, G2 in
all models). This should not be a serious cause for concern, because
the variance in the number of massive subhaloes hosting these luminous
satellites is very large and depends on the detailed merger history of
the halo. Moreover, the predicted number in the SAM is very sensitive
to the random orbit chosen, as discussed in
Section~\ref{ssec:cfrnbody}.  To assess this issue in a more robust
way, we need to increase the number of merger tree realizations of MW
mass haloes. We do this using semi-analytic merger trees instead of
the merger trees extracted from N-body simulations.  For each model we
generated 40 merger trees for DM haloes in the estimated mass range of
the Milky Way dark matter halo, $(0.8-1.2)$ $\times 10^{12}\hMsun$
(Klypin, Zhao \& Somerville 2002).  Each of the SAM codes has its own
algorithm for generating merger trees: the K08 and S08 model use
different implementation of the EPS algorithm, while \morgana~ uses
the {\sc pinocchio} code. Figure \ref{fig:AveMv} shows the averaged
luminosity function for the three semi-analytic models (we tested that
the sub-halo mass function from the EPS/{\sc pinocchio} trees is in
agreement with the one extracted from the Nbody simulations).

The K08 and S08 model are, again, in good agreement with the
observational data: they are able to fit the Milky Way luminosity
function in the range $-15< M_V<-2$ (KS test results: 90.9\% for S08 and
90.5\% for K08). 
The K08 model shows a small
deficit of satellites at brighter magnitudes (especially if compared
with S08 and \morgana) but this occurs where the number of observed
satellites has a large error bar due to poor number statistics.  Our
models produce much better agreement with the number of luminous
satellites with $M_V<-15$ relative to the predictions of Benson \etal
(2002). Because of the large number of differences between the Benson
\etal models and those considered here, we can only speculate on the
source of this difference.

It is also interesting to note that the K08 and S08 models suggest that
the LF of ultra-faint satellites has a downward kink below about $M_V
\sim -6$, in better agreement with the SK08 observational results
adopting an NFW, rather than isothermal, radial density distribution
for the satellites. An NFW distribution for satellites is predicted by
hydrodynamic simulations (Macci\`o \etal 2006).  The comparison
between the \morgana~ model and the observational results can only be
performed down to $M_V=-5$ in this case. This because the {\sc
  pinocchio} code has never been tested on such small scales (e.g. Li
\etal 2006) and we did not feel confident in using merger trees with a
mass resolution below $\sim 10^8 \hMsun$. In the tested range for
$M_V$ the \morgana~ model is also in good agreement with observations
(KS test: 90.2\% ).

%--------------------------------------------------------------------------
%--------------------------------------------------------------------------

\begin{figure}
\psfig{figure=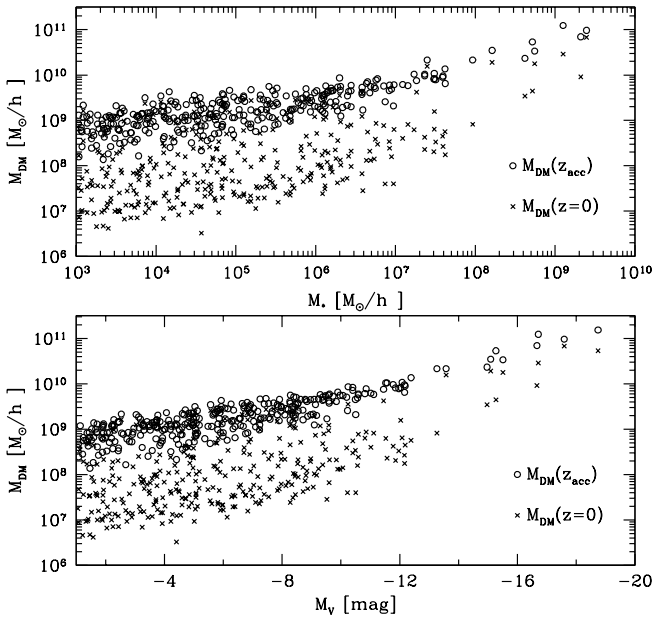,width=0.5\textwidth}
\caption{\scriptsize The dark matter mass of galaxy satellites versus
  their luminosity (lower panel) and stellar mass (upper panel). Open
  circles show the masses at time of accretion, while crosses show the
  present (z=0) dark matter mass.  Results are shown for the K08 model
  for all haloes G0-G3.}
\label{fig:Mdm}
\end{figure}

In figure \ref{fig:Mdm} (lower panel) we plot the stellar mass and
luminosity of galactic satellites versus their dark matter subhalo
mass; results are shown for the K08 model for all haloes, and are
similar for the other models.  For this comparison we used both the
present dark matter mass ($M_{DM}(z=0)$) and the mass of the subhalo
at the time of accretion ($M_{DM}(z_{acc})$). The difference between
the two reflects the effects of tidal stripping on the dark matter
substructure. The correlation
between the present day dark matter mass and luminosity is quite broad
and, for low luminosities, $M_{DM}(z=0)$ at a fixed luminosity spans
almost 3 orders of magnitude.  This is because tidal stripping of the
dark matter subhalo washes out the initial correlation between
luminosity and $M_{DM}(z_{acc})$. The same applies to the comparison
between the dark and stellar mass of galactic satellites, as shown in
the upper panel of figure~\ref{fig:Mdm}.

Recently, Strigari \etal (2008) pointed out that one of the curious
properties of the newly discovered population of faint
satellites is that over four orders of magnitude in luminosity, these
objects seem to contain a nearly constant total mass within a radius
of 300 pc. If we focus on the faint population in
figure~\ref{fig:Mdm}, with $-3 < M_V < -10$, we can see that the huge
scatter in DM mass at fixed luminosity or stellar mass provides a
partial explanation for this apparent ``common'' mass scale for the
faint satellites. Macci\`o, Kang \& Moore (2009, see also Li \etal 2009) investigated
this in more detail, and presented a direct comparison between the
predicted luminosity and the mass within 300 pc for faint
satellites in our simulations. They argued that the inner profiles of
haloes that are initially very concentrated are less effected by tidal
heating than haloes that are less concentrated, so that the mass
within 300 pc is reduced for more massive subhaloes (which are less
concentrated) relative to less massive ones. When they corrected for
this concentration-dependent modification of the inner density
profile, they found that the Strigari \etal (2008) results are
quantitatively reproduced by our simulations.

%----------------------------------------------------------------------------
%----------------------------------------------------------------------------

\subsection{Luminosity Function in the Observational Plane}
\label{ssec:SDSS}

\begin{figure}
\psfig{figure=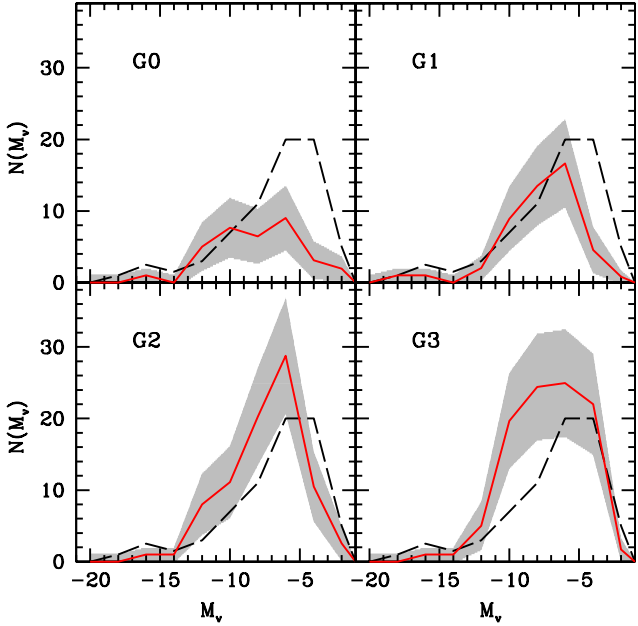,width=0.5\textwidth}
\caption{\scriptsize Comparison between the ``raw'' SDSS data
  (uncorrected for completeness) and model predictions with the SDSS
  visibility criteria applied to the simulations (see text for
  details).  The observational data are shown by the (black) dashed
  line, the K08 model is shown by the solid (red) line, and the shaded
  area shows the 1$\sigma$ Poisson scatter around the mean value.}
\label{fig:MVraw}
\end{figure}

The luminosity function of SK08 has been determined under certain
assumptions for the radial distribution of satellites around our
Galaxy. It is also interesting to apply the observational selection
criteria to our simulations and compare ``in the observational
plane'', i.e. with the {\it raw} data from the SDSS without
completeness corrections applied. To make such a comparison we have
applied to our sample of satellites a {\it visibility} criterion, in
order to determine if a given satellite would be detected in the SDSS.
We assume that {\em all} satellites brighter than $M_V=-10$ would be
visible and included in the SDSS sample. For fainter satellites, we
adopt a criterion based on both satellite distance and luminosity (see
SK08 for more details):
\begin{equation}
\log(R/{\rm kpc}) < 1.04 - 0.228 \times M_V.
\label{eq:vis}
\end{equation}
In the above formula the distance $R$ is measured from the Sun and not
from the center of the galaxy.  In order to convert our
galacto-centric distances into helio-centric distances, we assume the
Sun to be located at 8 kpc from the center of the galaxy (8,0,0);
moreover, since for each satellite galaxy we only know its distance
from the galactic center, we randomly assign position angles
$(\phi,\theta)$ to each of them (see also Tollerud \etal 2008).  We
exclude from the comparison galaxies more distant than 280 kpc.
Moreover since the SDSS covers only approximately one fifth of the sky
we randomly select 1/5 of our satellites.  We then average over 100
different realizations of this random sampling.  Observational data
for the recently discovered SDSS satellites are taken from MdJR08.
Figure~\ref{fig:MVraw} shows the comparison between the observations
and the luminosity functions obtained with the K08 model (the S08
model gives very similar results and this comparison is not possible
for the {\sc morgana} model, because it does not provide the distance
of satellite galaxies from the main halo).  The direct comparison with
the observational data confirms our previous results on the luminosity
function.  Our models are able to reproduce the data for haloes G0-G2,
while halo G3 slightly overproduces the number of faint
satellites. The agreement between the data and models implies that the
distance-luminosity relation of our satellites is similar to the
observed one.

\begin{figure}
\psfig{figure=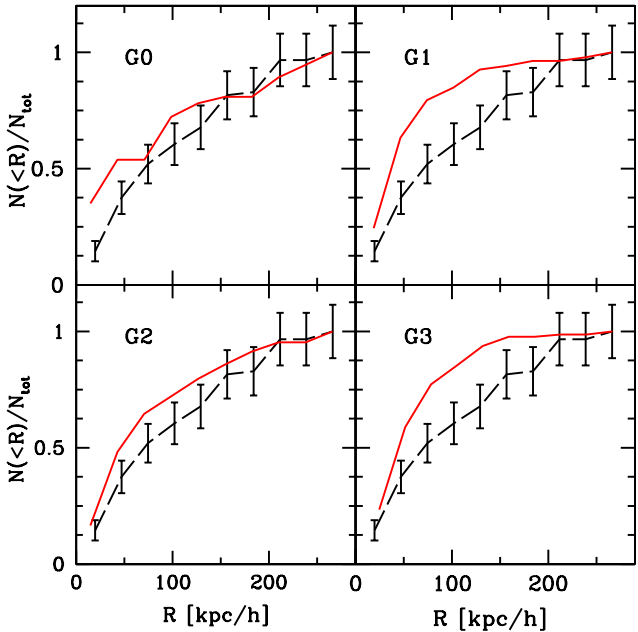,width=0.5\textwidth}
\caption{\scriptsize The fraction of satellites as a function of the
  distance from the central galaxy.  Observational data are shown by
  the (black) dashed line with error bars (representing the Poisson
  noise).  The K08 model is shown by the solid (red) line.}
\label{fig:Rraw}
\end{figure}

The application of the selection criteria of the SDSS to our simulated
data also allows us to compare the radial number density of satellites
in our models {\it vs } observations.  Results are shown in figure
\ref{fig:Rraw}. From this figure we see that our simulations reproduce
both the observed slope and normalization of the satellite radial
distribution for G0 and G2 (KS test results: 96\% and 91\%)
while the agreement is less good for G1 and G3 (KS: 76\% and 77\%).
The difference for G3 can be ascribed to the overall higher visible 
satellite number (e.g. fig. \ref{fig:MvAll}). One possible 
explanation for G1 can be related to its higher formation redshift (half mass in place) 
than the other two galaxies. This implies that subhaloes will have, on average
a higher accretion redshift, and thus have more time to sink to the center.

%%%%%%%%%%%%%%%%%%%%%%%%%%%%%%%%%%%%%%%%%%%%%%%%%%%%%%%%%%%%%%%%%%%%%%

%%%%%%%%%%%%%%%%%%%%%%%%%%%%%%%%%%%%%%%%%%%%%%%%%%%%%%%%%%%%%%%%%%%%%%

\begin{figure}
\psfig{figure=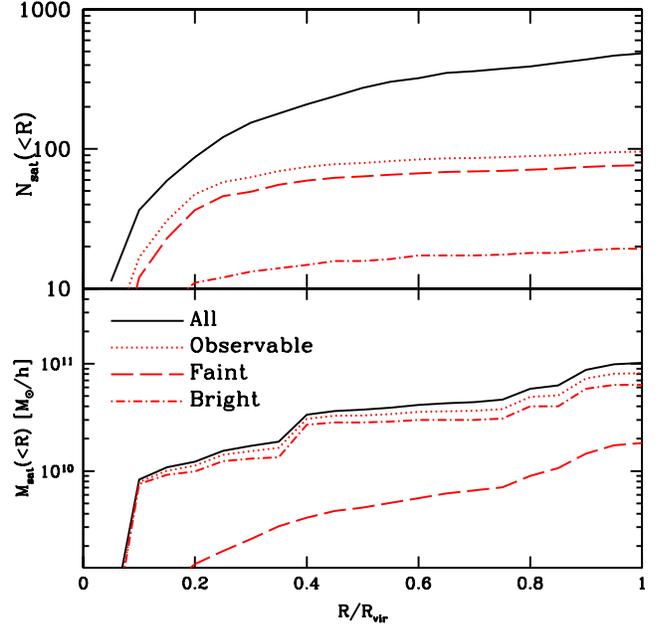,width=0.5\textwidth}
\caption{\scriptsize Upper panels: number-weighted cumulative
  fractional radial distribution of satellites; the black line shows
  all substructures (dark and observable), the (red) dotted, dashed
  and dot-dashed lines are for the ``observable'' satellites
  ($M_V<-1$), the faint satellites ($-9<M_V<-1$) and the
  classical satellites ($(M_V<-9$), respectively. Lower panel: same as
  upper panels but with weighted by satellite mass. The results were
  obtained by averaging over the four haloes G0-G3; only the K08 model
  is shown.}
\label{fig:distMS}
\end{figure}

Finally in figure \ref{fig:distMS} we compare the radial distribution
of DM substructures and ``observable'' satellites in the K08 model
(obviously, observability depends on many factors, but in this context
we simply consider all satellites with $M_V<-1$ to be
``observable''). In the upper panel we show the number density radial
distribution of all ``observable'' satellites ($M_V<-1$), of
faint satellites ($-9<M_V<-1$) and of classical satellites
($M_V<-9$). The number-weighted distribution of observable
satellites (which are dominated by the much more numerous faint
population) traces the sub-halo distribution (but is down by a factor
of $\sim 2$) at small radii (within $R/\Rvir \approx 0.2$), but
flattens relative to the sub-halo distribution at large radii. This
implies that ``observable'' sub-haloes are more concentrated near the
large galaxy than the overall population of subhaloes (see also
Kravtsov et al. 2004). When satellites are weighted by their mass
(lower panel of the same figure), the distribution is dominated by the
classical (bright) satellites, and is almost identical to the
mass-weighted distribution for all sub-haloes. Thus, the different
radial distribution of ``observable'' and dark satellites is due to
the suppression of star formation in low-mass haloes due to cosmic
reionization and feedback processes.

%-------------------------------------------------------------------------------------------------------

%-------------------------------------------------------------------------------------------------------

\subsection{Physical Processes that Shape the Satellite Luminosity Function}
\label{ssec:phys}

We have shown that it is indeed possible to reproduce the observed
Milky Way satellite luminosity function within the \LCDM model; we now
investigate the role of various physical processes in shaping the LF
in our theoretical models, with a focus on the origin of the newly
discovered ultra-faint satellite population. There are several
possible physical origins for the ultra-faint satellite population: it
can originate from (i) object that formed in haloes with $T<10^4$ K
via H$_2$ cooling (e.g. Salvadori \& Ferrara 2009); (ii) haloes with
$T>10^4$K that were inefficient at accreting (hot) gas, because of
photoionization; (iii) haloes with $T>10^4$K in which star formation
was inefficient because of strong supernova feedback; (iv) objects
that originally had larger stellar masses but have experienced
significant stellar stripping.

In the following sections we will investigate scenarios (ii)-(iv). The
first scenario, in which the ultra-faint dwarfs form via H$_2$ cooling
in haloes with $T<10^4$ K, is not accounted for in our models. It is
likely however that some ultra-faint satellites could form in this
way, and this could help in explaining the small gap between the
theoretical predictions and observations for $M_V \gta -3$
(e.g. figure \ref{fig:AveMv}).

\subsubsection{Reionization}
\label{ssec:reion}

We test the effect of our adopted parametrization of cosmic
reionization on our results, by both varying the reionization redshift
within our reference model based on Kravtsov \etal (2004) and by
applying a simple modification to this model to take into account
recent results based on high resolution hydrodynamical simulations
(e.g. Okamoto \etal 2008; Hoeft \etal 2006).

\begin{figure}
\psfig{figure=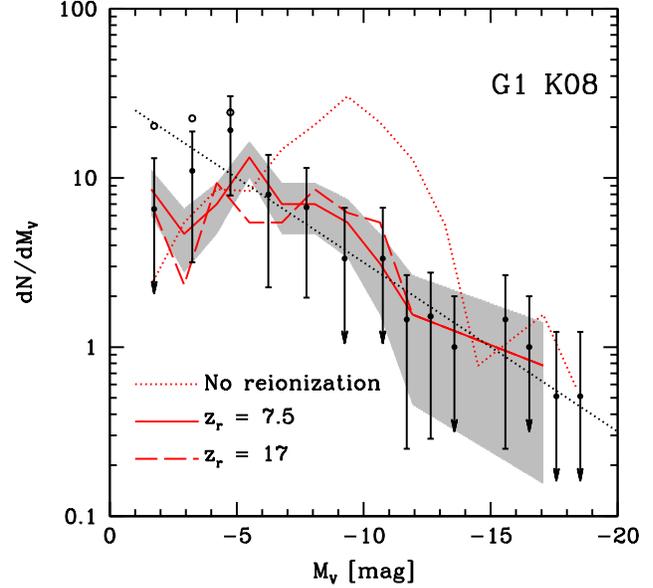,width=0.5\textwidth}
\caption{\scriptsize Satellite luminosity function of the G1 halo in
  the K08 model for three different reionization redshifts. The solid
  line represents our fiducial model with $z_r=7.5$, the dashed line
  is for $z_r=17$, and the dotted line is for a model with no
  reionization (but including all other kinds of feedback).  The
  shaded area represents the $1\sigma$ scatter around the mean of the
  $z_r=7.5$ model.}
\label{fig:reion}
\end{figure}

The redshift at which reionization occurs is still quite uncertain,
but it is bracketed in the range $7<z_r<15$ (3$\sigma$ range from
Komatsu \etal 2009); moreover, due to the fact that reionization
proceeds in an inhomogeneous way, the actual redshift of reionization
for the Local Group could substantially differ from the average
reionization redshift of the Universe (Weinmann \etal 2007).  Figure
\ref{fig:reion} shows the impact of varying the reionization redshift
on the G1 luminosity function for the K08 model using the standard
reionization parametrization (S08 and \morgana~ show a similar trend).
Without any suppression of gas accretion due to reionization, the
simulated LF contains too few satellites fainter than $M_V=-5$ and too
many with $M_V \sim -9$. This is because, in absence of reionization,
hot gas can cool very efficiently via atomic cooling, and every halo
can transform a large fraction of its gas content into stars before SN
feedback shuts star formation down.  When the effect of reionization
is taken into account, the amount of gas available for cooling and
star formation is reduced in low mass haloes, and many galaxies are
shifted from intermediate luminosities ($-15<M_V<-6$) to low
luminosities ($M_V>-6$), producing a luminosity function that is close
to a power-law and is in good agreement with the data. It is also
interesting to note that the LF is almost insensitive to the redshift
of reionization (solid line shows results for $z_r=7.5$, dashed line
for $z_r=17$); this is in agreement with earlier results obtained by
Kravtsov \etal (2004).

%----------------------------------------------------------------------------
%-----------------------------------------------------------------------------

\begin{figure}
\psfig{figure=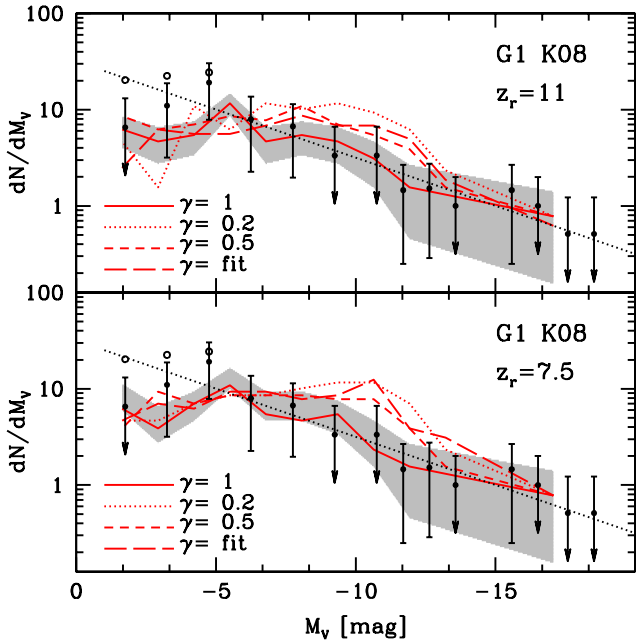,width=0.5\textwidth}
\caption{\scriptsize Satellite luminosity function for the G1 halo in
  the K08 model, for different parametrizations of cosmic
  reionization.  The solid line shows results for the standard
  reionization model ($\gamma = 1$, see text for the definition of $\gamma$)
  based on Kravtsov \etal (2004). 
  The dotted and short dashed lines show results for a
  filtering mass $M_F$ reduced by 80\% and 50\% respectively. The long
  dashed line is for a model with a redshift dependent expression for
  the modification of $M_F$ (see text for more details).  The shaded
  area represents the $1\sigma$ scatter around the mean of the
  $\gamma=1$ model.  The left upper and lower panels are for $z_r=11$ and
  $7.5$ respectively.}
 \label{fig:mvRT}
\end{figure}

In addition to uncertainty about the redshift of reionization, there
is still a debate about the value of the characteristic mass, $M_F$,
below which galaxies are strongly affected by photoionization. Okamoto
\etal (2008, see also Hoeft \etal 2006), using hydrodynamical
simulations, recently suggested that the actual value of $M_F$ can be
significantly lower than values previously obtained (e.g. by Gnedin
2000).  In order to explore the implications of their results, we
introduce a factor $\gamma$ that multiplies the original value of
$M_F(z)$ as derived by Gnedin (2000, see eq. \ref{eqn:fbar}).
We have used two constant
values for $\gamma$, namely 0.2 and 0.5, and a redshift dependent
expression $\gamma(z)=(1+z)^{1.1}/11.8$, derived from figure B1 of
Okamoto \etal (2008).  The results for these three modified models are
shown in figure \ref{fig:mvRT} for the G1 halo and for the K08 model
(results from the other models are similar). We see that the reduced
value of $M_F$ has the effect of increasing the number of galaxies
with $M_V\sim 10$, creating a bump in the LF, with respect to the
standard case, similar to what we saw in figure \ref{fig:reion} for
the no reionization run (dotted line).  This is true especially for
strong suppression of photo-ionization as in the $\gamma=0.2$ and
$\gamma(z)$ cases. Nevertheless it is still possible to reconcile the
simulated LFs with the observational data by increasing the
reionization redshift, as shown in the right panel of figure
\ref{fig:mvRT}, where we use $z_r=11$. 

In summary, it is interesting to note the interplay between the
strength of the suppression of gas accretion due to photo-ionization
(as reflected in the filtering mass scale) and the redshift of
reionization. In our standard models, in which the suppression is
relatively strong ($M_F$ is large), we find a weak dependence of the
predicted LF on the adopted value of the reionization redshift, while
in models with a lower overall normalization of the filtering mass, we
find a stronger dependence on the redshift of reionization.

%----------------------------------------------------------------------------
%---------------------------------------------------------------------------

\subsubsection{Stellar Stripping and Tidal Destruction}
\label{ssec:strip}

In our models, we distinguish between ``stellar stripping'' and
``tidal destruction''. In the former, we track the amount of {\em
 stellar} material that is stripped from the galaxy as it orbits
within the parent halo. In the latter, we assume that a satellite's
baryonic mass is unaffected until its total mass is stripped by a
critical amount, at which point the satellite is simply
removed. Obviously, these are aspects of the same physical process and
it is somewhat artificial to distinguish between them. We do so simply
to illustrate the sensitivity of our results to different
implementations of this physical process in models.

The \morgana~ model allows for the modelling of stellar stripping as a
satellite galaxy orbits around the parent galaxy (see section
\ref{sec:sam} for more details), while the K08 and S08 models assume
that the stellar content of a satellite remains unchanged until the
dark matter halo is stripped beyond a certain critical point, at which
point the galaxy is destroyed completely. 

\begin{figure}
\psfig{figure=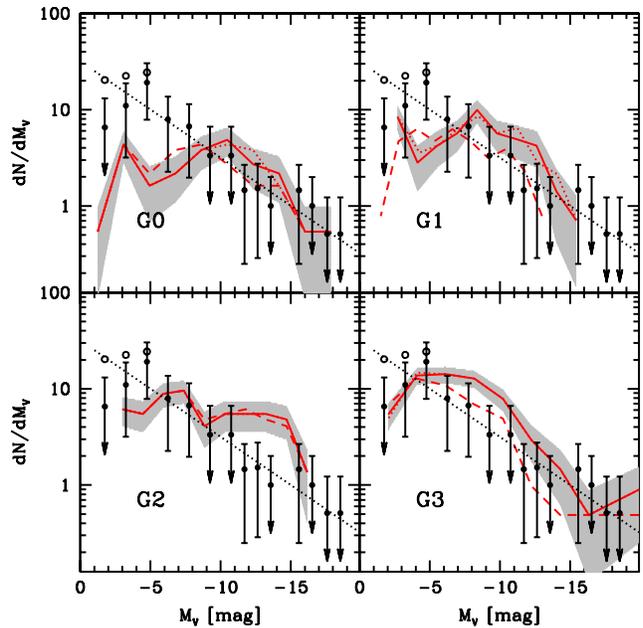,width=0.5\textwidth}
\caption{\scriptsize The effect of stellar stripping on the satellite
  luminosity function. Results from the \morgana~ model are shown for
  different levels of stellar stripping (no stripping, standard, and
  high). The solid line shows the moderate (standard) stripping case
  (as shown in figure \ref{fig:MvAll}); dotted and dashed lines show
  the no-stripping and strong stripping models respectively.}
\label{fig:stripmv}
\end{figure}

First we investigate the possible effect of stellar stripping on the
predicted satellite LF using the \morgana~ code. We compare three
different \morgana~ runs with no, standard (moderate) and high stellar
stripping; this latter case is obtained by increasing by a factor of
three the fraction of stripped stellar mass with respect to the
standard run (i.e. we remove from the satellites three times the mass
that is beyond the tidal radius).  Results for the luminosity function
are shown in figure \ref{fig:stripmv}.  In the case of standard
stripping the average stripped stellar mass is of the order $5\%-10\%$
with no mass dependence. Comparing the LF of this case with the
no-stripping case, it appears that stellar stripping is almost
negligible for satellites brighter that $M_V=-5$ and it marginally
affects fainter satellites.  When the strength of the stellar
stripping is increased the mass loss is, of course, more important and
as much as $40\%$ of the stellar mass can be stripped, with a strong
dependence on the orbital parameters. However, even in this case the
overall effect on the LF is relatively small.  According to these
results, stellar stripping is not one of the most important processes
that shapes the satellite LF or produces ultra-faint satellites.

%-----------------------------------------------------------------------------
%----------------------------------------------------------------------------

\begin{figure}
\psfig{figure=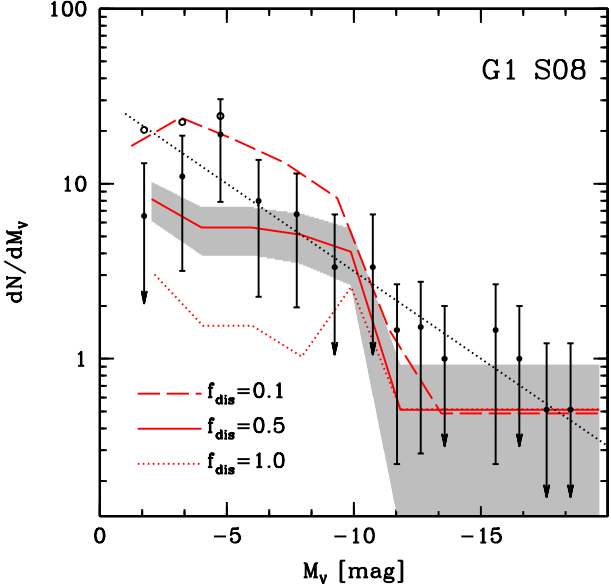,width=0.5\textwidth}
\caption{\scriptsize The effect of varying the tidal destruction
  parameter $f_{\rm dis}$ (see text) on the predicted MW satellite LF
  in the S08 model, for the G1 halo. 
}
\label{fig:S08fd}
\end{figure}

On the other hand we find that the modelling of tidal destruction of
satellites does have a significant effect on the satellite LF. If we
neglect tidal destruction, we find many more low-mass subhaloes than
are seen in the N-body simulations, and we would also predict many
more faint satellites than are observed in the MW.  Following Zentner
\& Bullock (2003) and Taylor \& Babul (2004), the S08 model considers
a satellite to be tidally destroyed when the mass of the dark matter
sub-halo has been stripped to a value less than or equal to the mass
within $f_{\rm dis} r_s$, where $r_s$ is the halo's original NFW scale
radius. Zentner \& Bullock (2003) adopt $f_{\rm dis}=1$ while Taylor
\& Babul (2004) adopt $f_{\rm dis}=0.1$. Figure~\ref{fig:S08fd} shows
the effect of varying the parameter $f_{\rm dis}$ in the S08 model; we
see that this can change the number of faint satellites by as much as
a factor of ten. We find good agreement with the observed MW satellite
LF for $f_{\rm dis}\sim0.1-0.5$.

%-----------------------------------------------------------------------------
%----------------------------------------------------------------------------

\subsubsection{Supernova Feedback}
\label{ssec:snfb}

\begin{figure}
\psfig{figure=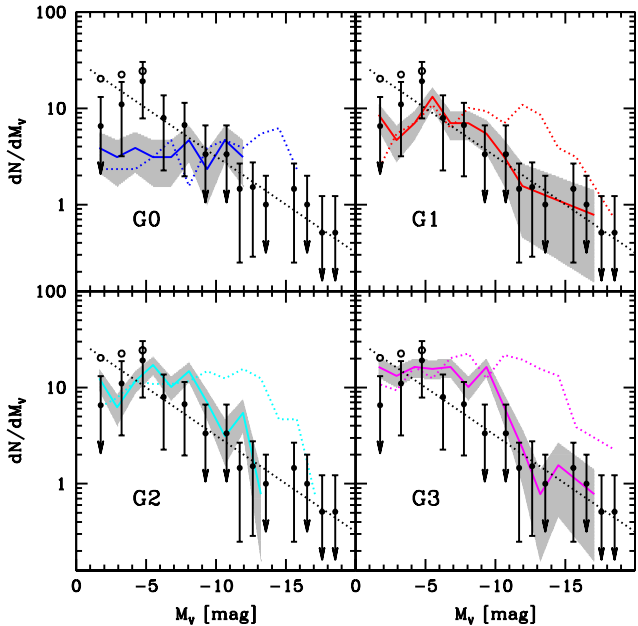,width=0.5\textwidth}
\caption{\scriptsize The effect of supernova feedback on the
  satellite luminosity function. Results from the K08 model with SN
  feedback switched {\it on} (solid line with shaded area) and {\it
    off} (dotted line) are shown. }
\label{fig:snfb}
\end{figure}

Feedback from supernovae is believed to be an important mechanism for
regulating star formation in low mass galaxies, and it plays a primary
role in shaping the LF in semi-analytic models. It is then important
to disentangle its effect from the effect of cosmic reionization
discussed in the previous section. To investigate this, we first
compare our reference model with both cosmic reionization and SN
feedback, and a run with only cosmic reionization (no SN
feedback). The results for the K08 model are shown in figure
\ref{fig:snfb}; the S08 model shows similar behaviour. In the absence
of stellar feedback, the SAM predicts a deficit of faint satellites
($M_V>-5$) and an overabundance of intermediate luminosity satellites
($-15<M_V<-10$) when compared with the observations. The inclusion of
SN-driven outflows again moves intermediate luminosity galaxies into
the ultra-faint regime by removing a significant fraction of the gas
and thereby suppressing star formation.

%-------------------------------------------------------------------------------------
%-------------------------------------------------------------------------------------

As we already discussed in Sec.~\ref{sec:sam} the original version
of the \morgana~ model, implementing a recipe for supernova
feedback derived from Monaco (2004), does not produce good agreement
with the MW satellite LF. In order to reproduce the observed MW
satellite LF, we found it necessary to  modify the stellar
feedback modeling using an approach similar to K08 and S08.

We conclude that strongly {\em differential} SN feedback (in
which the outflow rate relative to the star formation rate is much
higher in low-mass galaxies) plays a key role in reproducing the MW
satellite LF in these models. 

\subsubsection{Satellite Strangulation}

There is another effect that in our models can cause relatively
high-mass haloes ($T>10^4$K) to be inefficient at forming stars and
thus to produce ultra-faint satellites. This effect is a result of the
fairly standard assumption in SAMs of ``satellite strangulation'',
namely that the hot gas halo that could provide new gas to a galaxy is
stripped immediately when a galaxy becomes a satellite in a larger
halo. Thus, haloes that are accreted by the parent halo soon after
they crossed the threshold for atomic cooling ($T>10^4$K) and
thereafter are starved of any new gas cooling or accretion can have
very low stellar masses.

\begin{figure}
\psfig{figure=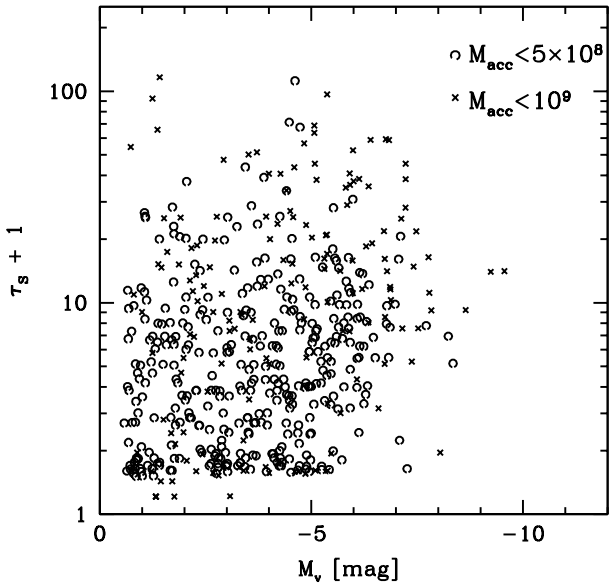,width=0.5\textwidth}
\caption{\scriptsize Relation between satellite luminosity and
  $\tau_{S}$ (defined according to eq.\ref{eq:tstrang}) for the K08
  model. Different symbols show different values for the total mass of
  the satellite at the time of accretion, as indicated in the figure
  label. Satellites with $\tau_{S} < 1$ may have been affected by
  ``strangulation''. }
\label{fig:tform2}
\end{figure}

In order to explore the importance of this effect, we look for a
correlation between satellite luminosity and the timescale $\tau_S$,
which we define as:
%----------------------------------------------------
\begin{equation}
\tau_{\rm S} = { {t_{\rm acc}-t_{\rm form}} \over {t_{\rm cool}}}
\label{eq:tstrang}
\end{equation}
%----------------------------------------------------
where the formation time ($t_{\rm form}$) is defined as the time at
which the halo reaches a virial temperature $T>10^4$K and can first
begin to cool, $t_{\rm acc}$ is the time at which the galaxy is
accreted by the parent halo and becomes a satellite, and the cooling
time ($t_{\rm cool}$) is the time needed for the gas to radiate away
all of its energy via atomic cooling, computed at $z=z_{acc}$ (the
standard definition of cooling time used in SAMs). Satellites with
$\tau_{S} < 1$ may have been ``strangulated'' before they were able to
cool a significant fraction of their gas. Figure \ref{fig:tform2}
shows that some (about 15 \%) of the faint satellites have $\tau_{S} <
1$ and therefore may have been impacted in part by this effect 
(though
photo-ionization and SN feedback may still play a role in these objects as
well). 

%-------------------------------------------------------------------------------------------------------
%-------------------------------------------------------------------------------------------------------

%%%%%%%%%%%%%%%%%%%%%%%%%%%%%%%%%%%%%%%%%%%%%%%%%%%%%%%%%%%%%%%%%%%%%%
%% SECTION 5: SUMMARY
%%%%%%%%%%%%%%%%%%%%%%%%%%%%%%%%%%%%%%%%%%%%%%%%%%%%%%%%%%%%%%%%%%%%%%
\section{Discussion and conclusions}
\label{sec:concl}

In the last few years, a new population of ultra-faint dwarf satellite
galaxies has been discovered around our Galaxy.  Given these new
observational data, it is timely to revisit the long standing problem
of the number of satellites around Milky Way-like dark matter haloes
as predicted in the \LCDM scenario.  We address this issue by
combining high resolution N-body simulations with three different
semi-analytic models of galaxy formation.  Four high resolution N-body
simulations are used to create detailed merger trees that represent
the assembly history of a Galactic dark matter halo. These merger
trees are then used as common input for three SAMs for galaxy
formation, namely the \morgana~ model (Monaco \etal 2007), the
Somerville \etal model, (S08, Somerville \etal 2008) and the Kang et
al. model (K08, Kang \etal 2008), 
to study the expected abundance and
properties of satellite galaxies in the Local Group.

Because the SAMs do not use the explicit information about subhaloes
from the N-body, but track subhalo evolution using semi-analytic
recipes, we first compare the mass function and radial distribution of
dark matter subhaloes predicted by the SAMs with the results directly
obtained from the N-body simulations. We find that the
parameterizations of subhalo merging and tidal stripping and
destruction adopted by the S08 and K08 models are able to fairly
accurately reproduce simulation results for the mass distribution and
radial distribution of satellites.

We then test the luminosity function of our simulated satellite
population against the latest observational results for the MW
satellite luminosity function. Our models are all able to reproduce
the LF down to a magnitude $M_V=-5$; at fainter magnitudes
($-5<M_V<-1$) the K08 and S08 models also provide a good fit to the
observational data, while the \morgana~ model tends to underestimate
the abundance of ultra-faint satellites, though the predictions are
still consistent with the MW data at the 1-$\sigma$ level.  All models
seem to suggest a decrease in the satellite number density below $M_V
\sim -5$, consistent with the assumption of a NFW like radial
distribution for observed satellites.

We also perform the comparison between our model predictions and the
observations in the ``observational plane'', i.e., by applying
``visibility'' criteria to the simulated satellites and comparing with
the SDSS data without any completeness corrections applied. In this
case, instead of assuming a radial distribution for the observed
satellites (which could in principle depend on e.g. satellite mass or
luminosity in a complex way), we make use of the predictions of our
models for the joint distribution function of satellite luminosity
and distance from the central galaxy. We again find good agreement,
increasing the robustness of our results.

We investigated the main physical processes responsible for shaping
the luminosity function of Milky Way satellites in our models. In the
absence of cooling by molecular hydrogen, and in the absence of
processes like photo-ionization, SN feedback, or stellar stripping,
the predicted satellite LF would show a peak at around $M_V \sim -14$
and a sharp drop-off at $M_V>-10$, with essentially {\em no} satellites
fainter than $M_V\sim-8$ predicted. This drop below $M_V>-8$ is {\rm
  not} due to limited numerical resolution, but rather to the sharp
assumed cooling cutoff at $T<10^4$K (because of our adopted atomic
cooling function). However, at temperatures just slightly above the
atomic cooling cutoff, cooling becomes quite rapid and so in the
absence of some kind of feedback or suppression mechanism, these
haloes rapidly cool all of the available baryons and convert them into
stars.

In our models, photo-ionization due to a cosmic reionizing background
and supernova feedback work together in order to re-shape this highly
peaked luminosity function into the near-power law down to $M_v \sim
-3$ that is implied by the recent SDSS observations. Photo-ionization
suppresses the infall of hot gas into low-mass dark matter haloes,
reducing the supply of baryons that are available for cooling and star
formation, while SN feedback reheats cold gas and expels it from
small haloes, again suppressing the efficiency of star formation. If
we include only photo-ionization or only SN feedback, we find an
excess of intermediate luminosity satellites and a shortage of
ultra-faint satellites (see also Koposov \etal 2009). In agreement
with previous works (e.g. Kravtsov \etal 2004), we find that the
satellite luminosity function in our ``standard'' models depends only
weakly on the assumed redshift of reionization.

We made use of results from the numerical hydrodynamic simulations of
Gnedin (2000) to motivate our treatment of the suppression of gas
infall due to the presence of a photo-ionizing background. A key
parameter in this recipe is the ``filtering mass'', or the halo mass
below which the gas content is significantly reduced relative to the
cosmic average. However, recent work by other groups (Okamoto \etal
2008; Hoeft \etal 2006) has found that the filtering mass may be
considerably smaller than the results of Gnedin (2000) suggested. We
investigate the implications of modifying the normalization and/or
redshift dependence of the filtering mass as suggested by these works,
and find that when a lower normalization of the filtering mass is
adopted, the results are more sensitive to the redshift of
reionization. We find that we can still reproduce the observed
luminosity function with the lower filtering mass if we adopt a higher
reionization redshift ($z_r \sim 11$ instead of $z_r \sim 7.5$).

We investigate the impact of stellar stripping on the observed
luminosity function using the \morgana~ model. We find that stellar
stripping can only decrease the satellite stellar masses by at most
about $\approx 20\%$, and therefore probably has only a minor effect on the
satellite LF. However, we find that the modelling of tidal {\em destruction} of
satellites does have a significant effect on faint end of the
predicted LF.

In this work, we have concentrated on the comparison with observations
of one particular quantity, the statistical distributions of 
satellite luminosity, although naturally the SAMs provide predictions of
many other galaxy properties (e.g. Lagos \etal 2009).
This has been done mainly for two
reasons: first because robust state of the art observational data have
recently been made available for those two quantities and second
because our goal was to directly address the so called missing
satellite problem and the origin of the newly discovered ultra-faint
population. Other observed properties  (such as metallicity, gas content, 
star formation history, etc.) contain complementary information 
on the formation mechanism of galactic satellites and 
more can be learned by studying in them in details
(e.g. Okamoto \etal 2009). While the objects produced 
in our SAMs do resemble several properties of observed 
satellites we have decided to defer a more extensive explorations 
of those properties to a future work.

The semi-analytic models that we use in this work were originally
normalized to reproduce global quantities such as the field galaxy
luminosity function or gas fractions for relatively luminous galaxies
($M_V\lta -16$). They have not previously been extensively tested
against observations of galaxies on these very small mass scales. In
the case of the K08 and S08 models, we found that the {\em identical}
model ingredients and even parameter values used in the standard
versions of these SAMs (e.g. Somerville \etal 2008, Kang \etal 2005)
were also able to reproduce the MW satellite LF. In the case of the
\morgana~ model, we found that the SN feedback recipe in the original
model (Monaco \etal 2007) had to be modified in order to reproduce the
faint satellite population. In either case, we have gained important
insights about the physical recipes incorporated in these models and
their applicability over a wide range of galaxy mass scales.

Our study confirms and expands on previous works that address the
so-called missing satellite problem using semi-analytic models and
simulations. The main new contribution of our paper is the
implementation of SAMs within merger trees extracted from numerical
simulations with very high mass resolution (particle mass $m_p\sim 4
\times 10^5 \hMsun$), in which each of our four Galaxy-sized halo
simulations contains 2--4 million particles. Thus, unlike previous
studies (e.g. Benson \etal 2002; Somerville 2002; Kravtsov \etal 2004)
which compared only with the ``classical'' satellite population
$M_V=-9$, we can resolve the very small subhaloes that may host the
newly detected population of ultra-faint satellites. We also have the
advantage of multiple simulations, unlike studies based on the Via
Lactea simulation (Diemand \etal 2007) alone. Although our results are
in qualitative agreement with studies based on simpler analytic
recipes for assigning baryons to dark matter (sub)-haloes
(e.g. Koposov \etal 2009), we showed that in our models, the MW
satellite LF is shaped by a complex combination of different physical
processes including tidal destruction, photo-ionization, and supernova
feedback.

Our models neglect several other physical processes that have been
discussed in the literature, and which may be important in shaping the
properties of galaxies on these mass scales. Although we model the
suppression of gas infall by a uniform cosmic radiation field after
reionization, we do not account for the modification of the atomic
cooling function by the radiation field, 
the possible photo-evaporation of gas from small haloes after
reionization (Barkana \& Loeb 1999), or photo-ionization by the nearby
large galaxy (i.e., the radiation field of the Milky Way;
e.g. Weinmann et al. 2007). Perhaps most importantly, we neglect
cooling via molecular hydrogen, and the associated complex and poorly
understood possible positive and negative feedback effects connected
with the formation and destruction of $H_2$ (e.g. Salvadori \& Ferrara 2009, Ricotti \etal 2008).
The fact that our models are nevertheless able to reproduce the bulk of the ultra-faint
satellite population {\em may} indicate that these other processes
operate at second order or cancel each other out, or may be simply a
fortuitous coincidence. Certainly this bears further study.

A further concern is that standard SAMs are known to fail to reproduce
the more detailed properties of satellites in larger mass hosts:
although they correctly reproduce the {\em number} of satellites as a
function of halo mass, several different SAM codes (including the ones
considered here) have been shown to produce too large a {\em fraction}
of red and passive satellites (e.g. Kimm \etal 2008 and references
therein) compared with observations.  The main cause of this
difficulty is believed to be the standard assumption that the hot gas
halo is immediately stripped when a satellite enters a larger host,
thereby depriving satellites of any new supply of gas 
(see Kang \& van den Bosch 2008 for a detailed discussion and possible
solution).  It is unclear whether this will impact our predictions for
the very low-mass MW satellites --- we defer this question to a future
investigation.

If we accept this ``baryonic'' solution of the ``missing satellite
problem'', other interesting implications follow.  Our analysis
predicts that roughly 1/5 of the total number of subhaloes with a
present-day bound mass $M>2 \times 10^7 \Msun$ should be dark. In
order to properly test this picture, a signature of the presence of
these dark satellites is needed. One possibility is that they could be
detected via gravitational lensing (e.g. Metcalf \& Zhao 2002) since
those small subhaloes will act as perturbers of the lensing signal
coming from the main halo. Unfortunately recent results based on
numerical simulations have shown that perturbations in the lensing
potential induced by (dark) satellites are very small and unlikely to
explain the anomalous flux ratios of some multiple lensed QSOs
(Macci\`o \etal 2006; Macci\`o \& Miranda 2006). Another possibility
would be detection of $\gamma$-rays from dark matter annihilation, as
the presence of substructure boosts the $\gamma$-ray signal by a
factor of 4 to 15 relative to smooth galactic models (Diemand \etal
2008).  Finally, a third possible opportunity to detect the presence
of a significant dark population of subhaloes in the MW halo could
come from the signatures of the interaction of such a population with
the thin stellar streams in the MW halo ( e.g.
Odenkirchen \etal 2001, Ibata \etal 2002, Johnston \etal 2002, Grillmair \& Dionatos 2006).

In final summary, our results show that not only is there no longer a
``missing satellite problem'', but that well-known and well-motivated
astrophysical processes working within the \LCDM framework {\em
  naturally} predict the form of the observed MW satellite luminosity
function over six orders of magnitude in luminosity. Indeed, it may be
that convincing proof of the existence of the large predicted
population of dark subhaloes via one of the methods suggested above
(or one not yet discovered) is one of the last remaining major
challenges for the \LCDM paradigm.

\section*{Acknowledgements} 

The authors are grateful to Jelte de Jong, Anna Gallazzi, Nicholas
Martin, Christian Maulbetsch, Hans-Walter Rix and Frank van den Bosch
for many stimulating discussions.  AVM thanks A. Knebe for his help 
with the {\sc AHF} halofinder. Numerical simulations were
performed on the PIA and on PanStarrs2 clusters of the
Max-Planck-Institut f\"ur Astronomie at the Rechenzentrum in Garching
and on the zBox2 supercomputer at the University of Z\"urich.  Special
thanks to B. Moore, D. Potter and J. Stadel for bringing zBox2 to
life. FF and SK acknowledge the Kavli Institute for Theoretical
Physics in Santa Barbara for hospitality: this research was partially
supported by the National Science Foundation under Grant No. NSF
PHY05-51164.  SK was supported by the DFG through SFB 439 and by a
EARA-EST Marie Curie Visiting fellowship.

%%%%%%%%%%%%%%%%%%%%%%%%%%%%%%%%%%%%%%%%%%%%%%%%%%%%%%%%%%%%%%%%%%%%%%
%%  REFERENCES
%%%%%%%%%%%%%%%%%%%%%%%%%%%%%%%%%%%%%%%%%%%%%%%%%%%%%%%%%%%%%%%%%%%%%% 


\begin{thebibliography}{}

%
\bibitem[Adelman-McCarthy et al.(2008)]{2008ApJS..175..297A} 
Adelman-McCarthy, J.~K., et al.\ 2008, \apjs, 175, 297 

%
\bibitem[Babul \& Rees(1992)]{1992MNRAS.255..346B} Babul, A., \& Rees, M.~J.\ 1992, \mnras, 255, 346 

\bibitem[Barkana \& Loeb (1999)]{1999ApJ...523...54B}
Barkana, R., Loeb, A., 1999, ApJ, 523, 54

%
\bibitem[Baugh(2006)]{2006RPPh...69.3101B} Baugh, C.~M.\ 2006, Reports on 
Progress in Physics, 69, 3101 

%
\bibitem[Belokurov et al.(2007)]{2007ApJ...654..897B} Belokurov, V., et 
al.\ 2007, \apj, 654, 897 

\bibitem[Benson et al.(2002)]{2002MNRAS.333..177B} Benson, A.~J., Frenk, 
C.~S., Lacey, C.~G., Baugh, C.~M., \& Cole, S.\ 2002, \mnras, 333, 177 

%*
\bibitem[Bertschinger(2001)]{2001ApJS..137....1B} Bertschinger, E.\ 2001, 
ApJS, 137, 1 

%*
\bibitem[Bullock, Kravtsov, \& Weinberg 2000]{2000ApJ...539..517B} Bullock J.~S., Kravtsov A.~V., 
Weinberg D.~H., 2000, ApJ, 539, 517 

%*
\bibitem[Chabrier (2003)]{2003PASP..115..763C}  Chabrier G., 2003, PASP, 115, 763 

%*
\bibitem[Diemand et al.(2007)]{2007ApJ...667..859D} Diemand, J., Kuhlen, M., \& Madau, P.\ 2007, \apj, 667, 859 

%*
\bibitem[Diemand et al.(2008)]{2008Natur.454..735D} Diemand, J., Kuhlen, 
M., Madau, P., Zemp, M., Moore, B., Potter, D., 
\& Stadel, J.\ 2008, Nature, 454, 735 

%*
\bibitem[Efstathiou(1992)]{1992MNRAS.256P..43E} Efstathiou, G.\ 1992, 
\mnras, 256, 43P 

\bibitem[Fontanot et al.(2009)]{2009MNRAS.397.1776F} Fontanot, F., De 
Lucia, G., Monaco, P., Somerville, R.~S., 
\& Santini, P.\ 2009, \mnras, 397, 1776 

\bibitem[Gilmore et al.(2007)]{2007ApJ...663..948G} Gilmore, G., Wilkinson, 
M.~I., Wyse, R.~F.~G., Kleyna, J.~T., Koch, A., Evans, N.~W., 
\& Grebel, E.~K.\ 2007, \apj, 663, 948 

\bibitem[Grillmair \& Dionatos(2006)]{2006ApJ...643L..17G} Grillmair, C.~J., \& Dionatos, O.\
2006, \apjl, 643, L17 

\bibitem[Hoeft et al.(2006)]{2006MNRAS.371..401H} Hoeft, M., Yepes, G., 
Gottl{\"o}ber, S., \& Springel, V.\ 2006, \mnras, 371, 401 

\bibitem[Ibata et al.(2002)]{2002MNRAS.332..915I} Ibata, R.~A., Lewis, 
G.~F., Irwin, M.~J., \& Quinn, T.\ 2002, \mnras, 332, 915 

\bibitem[Irwin et al.(2007)]{2007ApJ...656L..13I} Irwin, M.~J., et al.\ 2007, \apjl, 656, L13 

\bibitem[Johnston et al.(2002)]{2002ApJ...570..656J} Johnston, K.~V., 
Spergel, D.~N., \& Haydn, C.\ 2002, \apj, 570, 656 

\bibitem[Kang et al.(2005)]{2005ApJ...631...21K} Kang, X., Jing, Y.~P., Mo, 
H.~J., B\"orner, G.\ 2005, \apj, 631, 21 


\bibitem[Kang(2008)]{2008arXiv0806.3279K} Kang, X.\ 2008, Proceedings of IAU 254 "The Galaxy Disk in Cosmological Context", arXiv:0806.3279 

\bibitem[Kang \& van den Bosch(2008)]{2008ApJ...676L.101K} Kang, X., \& van den Bosch, F.~C.\ 2008, \apjl, 676, L101 

\bibitem[Kimm et al.(2009)]{2009MNRAS.394.1131K} Kimm, T., et al.\ 2009, 
\mnras, 394, 1131 

\bibitem[Klypin et al.(1999)]{1999ApJ...522...82K} Klypin, A., Kravtsov, 
A.~V., Valenzuela, O., \& Prada, F.\ 1999, \apj, 522, 82 

\bibitem[Klypin et al.(2002)]{2002ApJ...573..597K} Klypin, A., Zhao, H., 
\& Somerville, R.~S.\ 2002, \apj, 573, 597

\bibitem[Knollmann \& Knebe(2009)]{2009ApJS..182..608K} Knollmann, S.~R., \& Knebe, A.\ 2009, \apjs, 182, 608 

%*
\bibitem[Komatsu et al.(2009)]{2009ApJS..180..330K} Komatsu, E., et al.\ 
2009, \apjs, 180, 330 

\bibitem[Koposov et al.(2008)]{2007arXiv0706.2687K} Koposov, S., \etal 2008, ApJ, 686, 279 (SK08)

\bibitem[Koposov et al.(2009)]{2009ApJ...696.2179K} Koposov, S.~E., Yoo, 
J., Rix, H.-W., Weinberg, D.~H., Macci{\`o}, A.~V., 
\& Escud{\'e}, J.~M.\ 2009, \apj, 696, 2179 

\bibitem[Kravtsov et al.(2004)]{2004ApJ...609..482K} Kravtsov, A.~V., 
Gnedin, O.~Y., \& Klypin, A.~A.\ 2004, \apj, 609, 482 

\bibitem[Lagos et al.(2009)]{2009MNRAS.397L..31L} Lagos, C.~D.~P., Padilla, 
N.~D., \& Cora, S.~A.\ 2009, \mnras, 397, L31 

\bibitem[Li et al.(2007)]{2007MNRAS.379..689L} Li, Y., Mo, H.~J., van den 
Bosch, F.~C., \& Lin, W.~P.\ 2007, \mnras, 379, 689 

\bibitem[Li et al.(2009)]{2009MNRAS.397L..87L} Li, Y.-S., Helmi, A., De 
Lucia, G., \& Stoehr, F.\ 2009, \mnras, 397, L87 

\bibitem[Lo Faro et al.(2009)]{2009MNRAS.tmp.1225L} Lo Faro, B., Monaco, 
P., Vanzella, E., Fontanot, F., Silva, L., 
\& Cristiani, S.\ 2009, \mnras, 1225 

\bibitem[Macci{\`o} et al.(2006)]{2006MNRAS.366.1529M} Macci{\`o}, A.~V., 
Moore, B., Stadel, J., \& Diemand, J.\ 2006, \mnras, 366, 1529 

\bibitem[Macci{\`o} \& Miranda(2006)]{2006MNRAS.368..599M} 
Macci{\`o}, A.~V., \& Miranda, M.\ 2006, \mnras, 368, 599 


\bibitem[Macci{\`o} et al.(2007)]{2007MNRAS.378...55M} Macci{\`o}, A.~V., 
Dutton, A.~A., van den Bosch, F.~C., Moore, B., Potter, D., \& Stadel, J.\ 
2007, MNRAS, 378, 55 

\bibitem[Macci{\`o} et al.(2008)]{2008MNRAS.391.1940M} Macci{\`o}, A.~V., 
Dutton, A.~A., \& van den Bosch, F.~C.\ 2008, \mnras, 391, 1940 

\bibitem[Macci{\`o} et al.(2008)]{2008arXiv0810.1734M} Macci{\`o}, A.~V., Kang, 
X., \& Moore, B.\ 2009, ApJL, 692, L109

\bibitem[Madau et al.(2008)]{2008ApJ...689L..41M} Madau, P., Kuhlen, M., 
Diemand, J., Moore, B., Zemp, M., Potter, D., 
\& Stadel, J.\ 2008, \apjl, 689, L41 

%*
\bibitem[Mainini et al.(2003)]{2003ApJ...599...24M} Mainini, R., 
Macci{\`o}, A.~V., Bonometto, S.~A., \& Klypin, A.\ 2003, ApJ, 599, 24 

%*
\bibitem[Martin et al.(2007)]{2007MNRAS.380..281M} Martin, N.~F., Ibata, 
R.~A., Chapman, S.~C., Irwin, M., \& Lewis, G.~F.\ 2007, \mnras, 380, 281 

%*
\bibitem[Martin et al.(2008)]{2008ApJ...684.1075M} 
Martin, N.~F., de Jong, J.~T.~A., \& Rix, H.-W.\ 2008, \apj, 684, 1075  (MdJR08)

%*
\bibitem[Mateo(1998)]{1998ARAA..36..435M} Mateo, M.~L.\ 1998, ARA\&A, 36, 435 

\bibitem[McConnachie et al.(2008)]{2008ApJ...688.1009M} McConnachie, A.~W., 
et al.\ 2008, \apj, 688, 1009 

%*
\bibitem[Metcalf \& Zhao(2002)]{2002ApJ...567L...5M} Metcalf, R.~B., \& Zhao, H.\ 2002, \apjl, 567, L5 

%*
\bibitem[Metz et al.(2007)]{2007MNRAS.374.1125M} Metz, M., Kroupa, P., 
\& Jerjen, H.\ 2007, \mnras, 374, 1125 

%*
\bibitem[Monaco et al.(2002)]{2002MNRAS.331..587M} Monaco, P., Theuns, T., 
\& Taffoni, G.\ 2002, \mnras, 331, 587 

\bibitem[Monaco(2004)]{2004MNRAS.352..181M} Monaco, P.\ 2004, \mnras, 352, 
181 

%*
\bibitem[Monaco et al.(2007)]{2007MNRAS.375.1189M} Monaco, P., Fontanot, 
F., \& Taffoni, G.\ 2007, \mnras, 375, 1189 

%*
\bibitem[Moore et al.(1999)]{1999ApJ...524L..19M} Moore, B., Ghigna, S., 
Governato, F., Lake, G., Quinn, T., Stadel, J., 
\& Tozzi, P.\ 1999, \apjl, 524, L19 

%*
%A Universal Density Profile from Hierarchical Clustering
\bibitem[Navarro  et  al.(1997)]{1997ApJ...490..493N} Navarro,  J.~F.,Frenk, C.~S., \& White, S.~D.~M.\ 1997, ApJ, 490, 493


\bibitem[Odenkirchen et al.(2001)]{2001ApJ...548L.165O} Odenkirchen, M., et 
al.\ 2001, \apjl, 548, L165 

%*
\bibitem[Okamoto et al.(2008)]{2008MNRAS.390..920O} Okamoto, T., Gao, L., 
\& Theuns, T.\ 2008, \mnras, 390, 920 

\bibitem[Okamoto et al.(2009)]{2009arXiv0909.0265O} Okamoto, T., Frenk, 
C.~S., Jenkins, A., \& Theuns, T.\ 2009, arXiv:0909.0265 

\bibitem[Parkinson et al.(2008)]{2008MNRAS.383..557P} Parkinson, H., Cole, 
S., \& Helly, J.\ 2008, \mnras, 383, 557 

%*
\bibitem[Pe{\~n}arrubia et al.(2008)]{2008ApJ...672..904P} Pe{\~n}arrubia, 
J., McConnachie, A.~W., \& Navarro, J.~F.\ 2008, \apj, 672, 904 

\bibitem[Press et al.(1992)]{1992nrca.book.....P} Press, W.~H., Teukolsky, 
S.~A., Vetterling, W.~T., 
\& Flannery, B.~P.\ 1992, Cambridge: University Press, |c1992, 2nd ed.,  

%
\bibitem[Quinn et al.(1996)]{1996MNRAS.278L..49Q} Quinn, T., Katz, N., 
\& Efstathiou, G.\ 1996, \mnras, 278, L49 

%
\bibitem[Read et al.(2006)]{2006MNRAS.371..885R} Read, J.~I., Pontzen, 
A.~P., \& Viel, M.\ 2006, \mnras, 371, 885 

%*
\bibitem[Ricotti, Gnedin, \& Shull 2002]{2002ApJ...575...49R} Ricotti M., Gnedin N.~Y., Shull J.~M., 2002, ApJ, 575, 49 

%*
\bibitem[Ricotti, Gnedin, \& Shull 2008]{2008ApJ...685...21R} Ricotti
  M., Gnedin N.~Y., Shull J.~M., 2008, ApJ, 685, 21

%*
\bibitem[Salvadori 
\& Ferrara(2009)]{2009MNRAS.395L...6S} Salvadori, S., \& Ferrara, A.\ 2009, \mnras, 395, L6 

%*
\bibitem[Simon \& Geha(2007)]{2007ApJ...670..313S} Simon, J.~D., \& Geha, M.\ 2007, \apj, 670, 313 

\bibitem[Somerville \& Kolatt (1999)]{1999MNRAS.305..1S}
  Somerville, R.~S., Kolatt, T.~S., \mnras, 1999, 305, 1 

\bibitem[Somerville \& Primack (1999)]{1999MNRAS.310.1087S}
  Somerville, R.~S., Primack, J.~R., \mnras, 1999, 310, 1087

\bibitem[Somerville et al.(2001)]{2001MNRAS.320..504S} Somerville, R.~S., 
Primack, J.~R., \& Faber, S.~M.\ 2001, \mnras, 320, 504 

%*
\bibitem[Somerville(2002)]{2002ApJ...572L..23S} Somerville, R.~S.\ 2002,  \apjl, 572, L23 

\bibitem[Somerville et al.(2008)]{2008MNRAS.391..481S} Somerville, R.~S., 
Hopkins, P.~F., Cox, T.~J., Robertson, B.~E., 
\& Hernquist, L.\ 2008, \mnras, 391, 481 

%*
%Cosmological N-body simulations and their analysis
\bibitem[Stadel(2001)]{2001PhDT........21S} Stadel, J.~G.\ 2001, 
Ph.D.~Thesis, University of Washington  

%*
\bibitem[Strigari et al.(2007)]{2007ApJ...669..676S} Strigari, L.~E., 
Bullock, J.~S., Kaplinghat, M., Diemand, J., Kuhlen, M., 
\& Madau, P.\ 2007, \apj, 669, 676 

\bibitem[Strigari et al.(2008)]{2008Natur.454.1096S} Strigari, L.~E., 
Bullock, J.~S., Kaplinghat, M., Simon, J.~D., Geha, M., Willman, B., 
\& Walker, M.~G.\ 2008, Nature, 454, 1096 

%*
\bibitem[Taylor \& Babul (2004)] {taylor.2004} Taylor, J. E., \&  Babul, A., 2004, MNRAS,
  348, 811

%*
\bibitem[Thoul 
\& Weinberg(1996)]{1996ApJ...465..608T} Thoul, A.~A., \& Weinberg, D.~H.\ 1996, \apj, 465, 608 

%*
\bibitem[Tollerud et al.(2008)]{2008ApJ...688..277T} Tollerud, E.~J., 
Bullock, J.~S., Strigari, L.~E., \& Willman, B.\ 2008, \apj, 688, 277 


%*
\bibitem[Walsh et al.(2009)]{2009AJ....137..450W} Walsh, S.~M., Willman, 
B., \& Jerjen, H.\ 2009, AJ, 137, 450 

\bibitem[Wechsler et  al.(2002)]{2002ApJ...568...52W} Wechsler, R.~H.,
Bullock, J.~S.,  Primack, J.~R., Kravtsov, A.~V., \&  Dekel, A.\ 2002,
ApJ, 568, 52

%*
\bibitem[Weinmann et al.(2007)]{2007MNRAS.381..367W} Weinmann, S.~M., 
Macci{\`o}, A.~V., Iliev, I.~T., Mellema, G., 
\& Moore, B.\ 2007, \mnras, 381, 367 

%*
\bibitem[Willman et al.(2005)]{2005ApJ...626L..85W} Willman, B., et al.\ 
2005, \apjl, 626, L85 

%*
\bibitem[Zentner \& Bullock (2003)]{2003ApJ...598...49Z} Zentner,
  A. R., Bullock, J. S. 2003, ApJ, 598, 49

%*
\bibitem[Zentner et al. (2005)] {2003ApJ...598...49K} Zentner, A. R., Berlind, A. A.,
  Bullock, J. S., Kravtsov, A. V., Wechsler, R. H., 2005, ApJ, 624,
  505

%*
\bibitem[Zucker et al.(2006)]{2006ApJ...643L.103Z} Zucker, D.~B., et al.\ 
2006a, \apjl, 643, L103 


\end{thebibliography}
\end{document}